\documentclass[a4paper,12pt]{article}

\usepackage{amsfonts}
\usepackage{amsmath,epsfig}
\numberwithin{equation}{section}
\begin{document}

\title{Circularly-polarized plane waves in a deformed Hadamard material}
\author{Michel Destrade, Michael Hayes}

\date{2002}
\maketitle

\bigskip

%

\begin{abstract}
Small amplitude inhomogeneous plane waves propagating in any  
direction in a homogeneously deformed Hadamard material are 
considered. 
Conditions for circular polarization are established.
The analysis relies on the use of complex vectors (or bivectors) to 
describe the slowness and the polarization of the waves.

Generally, homogeneous circularly-polarized plane waves may propagate 
in only two directions, the directions of the acoustic 
axes, in a homogeneously deformed Hadamard material.
For inhomogeneous circularly-polarized plane waves, 
the number of possibilities is far greater. 
They include an infinity of `transverse waves', as well as
`longitudinal waves', and the superposition of 
`transverse waves' and `longitudinal waves',
where `transverse' and `longitudinal' are used in the bivector sense.

Each and every possibility of circular polarization is examined in 
turn, and explicit examples of solutions are given in every case.
\end{abstract}

\newpage

\section{Introduction}
When a homogeneous, isotropic, compressible elastic body is maintained
in a state of  finite homogeneous static deformation,
longitudinal waves are, in general, only possible when
the propagation direction is along a principal axis of strain.
However, Hadamard \cite{Hada03} introduced a
remarkable class of elastic materials characterized by the property
that \textit{infinitesimal} longitudinal waves may propagate
in every direction,
irrespective of the basic static finite homogeneous deformation. 
Many studies of the properties of Hadamard materials have been made
\cite{John66,CuHa69,Ogde70}. 
Among these we mention a result of
Boulanger, Hayes, and Trimarco \cite{BoHT94},
who showed that there are only two directions, $\mathbf{n^+}$ and
$\mathbf{n^-}$, called `acoustic axes', along which
\textit{finite-amplitude} circularly-polarized transverse waves may
propagate.
The acoustic axes $\mathbf{n^\pm}$ are the
directions along the normals to the planes of central circular sections
of the ellipsoid $ \mathbf{x \cdot} \mathbb{B}^{-1}\mathbf{x}=1$,
where  the left Cauchy--Green tensor associated with the basic
deformation is denoted by $\mathbb{B}$.
The acoustic axes are determined solely by the basic static deformation
and are independent of the choice of material constants and of the
function which occurs in  the strain-energy function describing the
Hadamard material.

Here, consideration is restricted to the propagation of
\textit{infinitesimal} plane waves in a Hadamard material
maintained in a state of finite static homogeneous deformation,
and the primary emphasis of this paper is on
\textit{inhomogeneous} plane waves.
For these waves, the incremental displacement $\mathbf{u}$
is of the form
$\mathbf{u}=\mathbf{A} \exp{i \omega (N \mathbf{C\cdot x}-t)}$,
where $ \mathbf{A}$ and $\mathbf{C}$ are the amplitude and propagation
bivectors, respectively, $\omega$ is the real frequency, and  $N$
is the complex scalar slowness.
The propagation bivector $\mathbf{C} = \mathbf{C}^+ + i\mathbf{C}^-$
is prescribed with $\mathbf{C}^+ \ne \mathbf{0}$,
$\mathbf{C}^- \ne \mathbf{0}$,
and $\mathbf{A}$ and $N$ are sought such that $ \mathbf{u}$ satisfies
the equations of motion \cite{Haye84}.
This is equivalent to finding the eigenvalues $\rho N^{-2}$ and
eigenbivectors $\mathbf{A}$ of the acoustical tensor
$\mathbb{Q}(\mathbf{C})$:
$\mathbb{Q}(\mathbf{C})\mathbf{A}= \rho N^{-2} \mathbf{A}$.
Our special consideration is the determination of circularly-polarized
inhomogeneous plane wave solutions.
For these waves, the amplitude bivector $ \mathbf{A}$ must
be isotropic: $ \mathbf{A \cdot A}=0$.
We find  that there are two distinct sets of solutions, according as to
whether or not $ \mathbf{C}$ is chosen to be isotropic.

In the case where $\mathbf{C}$ is chosen nonisotropic
($\mathbf{C \cdot C} \ne 0$), the `projection tensor'
$\mbox{\boldmath $\Pi$}=
\mathbf{1} - \{ \mathbf{C} \otimes \mathbf{C}/(\mathbf{C \cdot C}) \}$,
may be introduced.
For it, $\mbox{\boldmath $\Pi$}^2=\mbox{\boldmath $\Pi$}$,
$ \mathbf{\Pi C}= \mathbf{0}$.
If $ \mathbf{n^+ \Pi n^+}=0$,  $ \mathbf{n^- \Pi n^-} \ne 0$,
then $\mathbf{\Pi n^+}$ is an isotropic amplitude bivector.
The orthogonal projection of the directional ellipse of $\mathbf{C}$,
onto the plane of central circular section
of the $\mathbb{B}^{-1}$-ellipsoid with normal $\mathbf{n^+}$,
is a circle.
There is an infinity of such choices of $\mathbf{C}$ and therefore,
an infinity of circularly-polarized inhomogeneous plane waves.
(Similar comments apply when $\mathbf{C \cdot C} \ne 0$,
 $ \mathbf{n^- \Pi n^-}=0$,  $ \mathbf{n^+ \Pi n^+} \ne 0$.)

There are only two bivectors $\mathbf{C}$ satisfying
$\mathbf{C \cdot C} \ne 0$,
 $ \mathbf{n^+ \Pi n^+}=0$,  $ \mathbf{n^- \Pi n^-} = 0$.
Then $\mathbf{\Pi n^+}$ and $\mathbf{\Pi n^-}$ are both isotropic
amplitude bivectors with the same eigenvalue.
There are \textit{two}
circularly-polarized inhomogeneous plane waves in this case.

Finally, if the orthogonal projections of the directional ellipse of
$\mathbf{C}$ upon either plane   of central circular section
of the $\mathbb{B}^{-1}$-ellipsoid is not a circle,
$ \mathbf{n^\pm \Pi n^\pm} \ne 0$, then the tensor
$\mbox{\boldmath $\Pi$} \mathbb{B}^{-1} \mbox{\boldmath $\Pi$}$
admits two orthogonal non-isotropic eigenbivectors
$\mathbf{A^+}, \mathbf{A^-}$, with different eigenvalues.
In this case, if the directional ellipse of $\mathbf{C}$ is chosen to
be similar and similarly situated to a (non-circular) central
elliptical section of a certain $\mathbb{M}$-ellipsoid,
$\mathbf{C \cdot} \mathbb{M} \mathbf{C}=0$, where $\mathbb{M}$
depends upon the finite deformation, then there is a corresponding
isotropic amplitude bivector $\mathbf{A}$ which is parallel to either
$ \{ \mathbf{C} / (\mathbf{C \cdot C})^{1/2} \} \pm i \mathbf{A^+}$
or $ \{ \mathbf{C} / (\mathbf{C \cdot C})^{1/2} \} \pm i \mathbf{A^-}$.
There is a corresponding infinity of inhomogeneous circularly-polarized
waves.

Turning now to the set of solutions for which $\mathbf{C}$ is
isotropic, $\mathbf{C \cdot C} = 0$, it is seen that
if $ \mathbf{C}$ is chosen isotropic  and also satisfies
$ \mathbf{C \cdot} \mathbb{B}^{-1} \mathbf{C}=0$, so that
the circle of $ \mathbf{C}$ lies in either plane of central circular
section of the $\mathbb{B}^{-1}$-ellipsoid, then the corresponding
amplitude bivectors $ \mathbf{A}$ are parallel to $ \mathbf{C}$.
There are two such waves.

If $ \mathbf{C}$ is isotropic, $\mathbf{C \cdot C} = 0$,
and if $\mathbf{C \cdot} \mathbb{B}^{-1}\mathbf{C} \ne 0$,
$\mathbf{C \cdot} \mbox{\boldmath $\Phi$} \mathbf{C}=0$,
where $ \mbox{\boldmath $\Phi$}$ is a certain real symmetric tensor which
depends only upon the basic  deformation,
then the circle of $ \mathbf{C}$ lies in a plane of central circular
section of the ellipsoid associated with $ \mbox{\boldmath $\Phi$}$.
Corresponding to either choice of $\mathbf{C}$, there are
two linearly independent isotropic eigenbivectors of
$\mathbb{Q}(\mathbf{C})$, each with the same eigenvalue.
Thus, with either choice of $ \mathbf{C}$, there are \textit{two}
circularly-polarized waves which may propagate.

Finally, for isotropic $\mathbf{C}$, if
 $\mathbf{C \cdot} \mathbb{B}^{-1} \mathbf{C}\ne 0$,
$\mathbf{C \cdot} \mbox{\boldmath $\Phi$} \mathbf{C} \ne 0$, all  isotropic
amplitude eigenbivectors are parallel to  $ \mathbf{C}$.
There is an infinity of such choices of $\mathbf{C}$, and therefore
also of circularly-polarized plane waves.

Examples are presented for every type of solution.

The paper is organized as follows.

In \S \ref{Basic equations} we recall the equations governing the
behavior of a Hadamard material subjected to a finite static
homogeneous deformation. For the constitutive equation, we follow
Boulanger \textit{et al} \cite{BoHT94} and choose to express the Cauchy
stress in terms of $\mathbf{1}$, $\mathbb{B}$, and $\mathbb{B}^{-1}$,
where $\mathbb{B}$ is the left Cauchy--Green strain tensor, rather than
in terms of $\mathbf{1}$, $\mathbb{B}$, and $\mathbb{B}^2$
\cite{CuHa69,Ogde70}.
Then we assume that an infinite body of Hadamard material is subjected
to a finite static pure homogeneous deformation.
We also introduce the `acoustic axes' in the deformed state, which
play an important role in the study of elastic waves.

Then, in \S \ref{Small-amplitude plane waves in a deformed Hadamard
material} we consider the propagation of small-amplitude inhomogeneous
plane waves in a deformed body of Hadamard material. The waves are of
complex exponential type and their associated amplitude and slowness
are described through the use of bivectors \cite{Haye84}. Explicitly,
the perturbation is the real part of $\mathbf{A} \exp i \omega N
(\mathbf{ C \cdot x}-t)$, where $\omega$ is the real frequency of the
wave, and $\mathbf{A}$, $\mathbf{C}$, and $N$ are complex quantities
called the `amplitude bivector', the `propagation bivector', and the
`complex scalar slowness', respectively. Incremental strain, strain
invariants, and stress are computed, leading to the derivation of the
equations of motion and the acoustical tensor.

Next, we seek circularly-polarized solutions, which correspond
\cite{BoHa93} to the iso\-tro\-py of $\mathbf{A}$ (that is $ \mathbf{ A
\cdot A}=0$), or equivalently, to a double eigenvalue of the acoustical
tensor.

The case of circularly-polarized waves
with a nonisotropic propagation bivector
$\mathbf{C}$ (that is $ \mathbf{ C \cdot C} \ne 0$) is treated in
\S \ref{waves with a non-isotropic bivector}.
We prove that the existence of such waves is determined by a condition
linking $\mathbf{C}$ to some tensors which depend only on the finite
static deformation.
Transverse ($ \mathbf{ A \cdot C}=0$) waves are found, as well as
waves which can be decomposed into the superposition of a transverse
($ \mathbf{ A \cdot C}=0$) wave and a longitudinal
($ \mathbf{ A \wedge C}= \mathbf{0}$) wave.

In \S \ref{waves with an isotropic bivector} we consider
circularly-polarized inhomogeneous plane waves with an isotropic
bivector $\mathbf{C}$ (that is $ \mathbf{ C \cdot C} =0$).
We show that longitudinal waves can propagate.
We also find all other circularly-polarized waves.

Finally, in \S \ref{Circularly-polarized homogeneous plane waves},
we specialize our results to the case of circularly-polarized
\textit{homogeneous} plane waves, when $\mathbf{C}$ is a real unit
vector in the direction of propagation.
The result established by Boulanger \textit{et al} \cite{BoHT94} for
finite-amplitude plane waves is recovered:
circular polarization for homogeneous waves  occurs only in the
directions orthogonal to either of the planes of
central circular sections of the ellipsoid
$\mathbf{x \cdot} \mathbb{B}^{-1} \mathbf{x}=1$, where $\mathbb{B}$
is the left Cauchy--Green strain tensor of the finite static
deformation.

\section{Basic equations}
\label{Basic equations}

\subsection{Hadamard materials}
\label{Hadamard materials}

We consider homogeneous isotropic hyperelastic materials of
the Hadamard type.
These are characterized by a strain-energy density $\Sigma$, measured
per unit volume of the undeformed state, given by \cite{John66}
\begin{equation} \label{sigma}
2 \Sigma= a II + b I + f(III),
\end{equation}
where $a$, $b$ are two material constants. Also, $f(III)$ is a material
function, and $I, II, III$ are principal invariants of the left
Cauchy--Green tensor $\mathbf{B}$:
\begin{equation}
I= \text{ tr } \mathbf{B}, \quad 2II= I^2 - \text{ tr } (\mathbf{B}^2),
\quad III= \text{ det } \mathbf{B},
\end{equation}
with $\mathbf{B}$ given by
\begin{equation}
\mathbf{B}= \mathbf{F F}^{\mathrm{T}}, \quad B_{ij}= (\partial x_i /
\partial X_A) (\partial x_j /\partial X_A), \quad F_{iA}= (\partial x_i /
\partial X_A),
\end{equation}
where $x_i$ are the coordinates at time $t$ of a particle whose
coordinates are $X_A$ in the undeformed state. The deformation gradient
$\mathbf{F}$ is such that $ \text{ det } \mathbf{F} > 0$.

The constitutive equation for the Cauchy stress $\mathbf{t}$ for a
Hadamard material is \cite{BoHT94}
\begin{equation} \label{t}
\mathbf{t} = [ a II \, III^{-1/2} + g(III^{1/2})]\mathbf{1}
+ b \mathbf{B} -a III \mathbf{B}^{-1},
\end{equation}
where the function $g$ is defined by
\begin{equation} \label{g}
g=g(III^{1/2})= III^{1/2} f'(III).
\end{equation}

We exclude consideration of the special case where $a=0$, when the
material is  a `restricted Hadamard material'
\cite{Haye68,Will77,ChJa79,Dest99}.
Then, assuming $a \ne 0$, it may be shown
\cite{Ogde70,BoHT94} that in order for the Strong Ellipticity
conditions to hold, the following inequalities must be valid:
\begin{equation} \label{S-E}
a > 0, \quad b \ge 0, \quad g' \ge 0.
\end{equation}
It is assumed throughout that these conditions are satisfied.

We also assume that the body of Hadamard material is free of stress
in the undeformed state.
Thus $\mathbf{t}= \mathbf{0}$ when $\mathbf{B}=\mathbf{1}$.
Then we must have \cite{BoHT94}
\begin{equation}
f'(1)=-(a+2b).
\end{equation}

The equations of motion, in the absence of body forces, are
\begin{equation} \label{Motion}
\rho \mathbf{\ddot{x}}= \text{ div } \mathbf{t}, \quad \rho \ddot{x}_i
= \partial t_{ij} / \partial x_j,
\end{equation}
where $ \ddot{x}_i$ are the acceleration components. Also, $\rho$ is
the current mass density, related to the density $\rho_0$ of the
material in the undeformed state through
\begin{equation}
\rho = III^{-1/2} \rho_0.
\end{equation}

\subsection{Homogeneously deformed Hadamard material}
\label{Homogeneous Deformation}

Now, we assume that a Hadamard  material undergoes a finite pure
 homogeneous static deformation, bringing a  particle initially at
$\mathbf{X}$ in the undeformed state to $\mathbf{x}$ in the deformed
state. Let ($O, \mathbf{i}, \mathbf{j}, \mathbf{k}$) be a rectangular
Cartesian coordinate system defined by an origin $O$ and the unit
vectors $\mathbf{i}, \mathbf{j}, \mathbf{k}$ along the three directions
of principal stretches. Then the finite homogeneous deformation is
given by
\begin{equation} \label{x}
\mathbf{x}= \lambda_1 X \mathbf{i} + \lambda_2 Y \mathbf{j} +
 \lambda_3 Z \mathbf{k},
\end{equation}
where $\lambda_\alpha$ ($\alpha = 1,2,3$) are the stretch ratios in
each principal direction.
Throughout the paper, it is assumed that these ratios are distinct and
ordered as
\begin{equation} \label{orderLambda}
 \lambda_1 > \lambda_2 > \lambda_3.
\end{equation}

Let  $\mathbb{F}$ and  $\mathbb{B}$ be the deformation gradient and
left Cauchy--Green strain tensor corresponding to this deformation.
Then
\begin{equation}
\mathbb{F}= \mathrm{diag} \, (\lambda_1, \lambda_2, \lambda_3)
, \quad
\mathbb{B}= \mathrm{diag} \, (\lambda_1^2, \lambda_2^2,
\lambda_3^2),
\end{equation}
and the corresponding strain invariants $I,II,III$ are
\begin{equation}
I= \lambda_{1}^2+ \lambda_{2}^2 +\lambda_{3}^2,
\quad
II= \lambda_{1}^2 \lambda_{2}^2 + \lambda_{2}^2  \lambda_{3}^2
+ \lambda_{1}^2 \lambda_{3}^2,
\quad
III= \lambda_{1}^2\lambda_{2}^2 \lambda_{3}^2.
\end{equation}

In this deformed state, two specific directions play an important
role with respect to the propagation of finite-amplitude
homogeneous plane waves.
They are the so-called `acoustic axes',
whose directions are the only ones along which finite-amplitude
circularly-polarized waves may propagate.
It has been proved \cite{BoHT94} for a Hadamard material
that the acoustic axes are the normals to the planes of central
circular sections of the $\mathbb{B}^{-1}$-ellipsoid
($\mathbf{x \cdot} \mathbb{B}^{-1}\mathbf{x} =1$),
which are along the unit vectors $\mathbf{n}^{\pm}$ defined by
\begin{align} \label{n+-}
& \mathbf{n}^{\pm} = \alpha \mathbf{i} \pm \gamma \mathbf{j},
    \quad \alpha^2 + \gamma^2 = 1,\nonumber \\
& \alpha = \sqrt{\frac{\lambda_2^{-2} - \lambda_1^{-2}}
            {\lambda_3^{-2} - \lambda_1^{-2}}},
    \quad
  \gamma = \sqrt{\frac{\lambda_3^{-2} - \lambda_2^{-2}}
        {\lambda_3^{-2} - \lambda_1^{-2}}}, \\
& \alpha^2 \lambda_3^{-2} + \gamma^2 \lambda_1^{-2} = \lambda_2^{-2},
    \quad
  \gamma^2 \lambda_1^2 + \alpha^2 \lambda_3^2 - \lambda_2^2 =
     \alpha^2 \gamma^2 III (\lambda_3^{-2} - \lambda_1^{-2})^2.
\nonumber
\end{align}

These unit vectors $\mathbf{n}^{\pm}$ also appear in the
Hamilton cyclic decomposition of the $\mathbb{B}^{-1}$ tensor
\cite{BoHa93},
\begin{equation} \label{Hamilton}
\mathbb{B}^{-1} = \lambda_2^{-2} \mathbf{1}
- \textstyle{\frac{1}{2}} (\lambda_3^{-2} - \lambda_1^{-2})
[ \mathbf{n}^+ \otimes \mathbf{n}^- +
    \mathbf{n}^- \otimes \mathbf{n}^+ ],
\end{equation}
where $\otimes$ denotes the dyadic product.

Finally, the constant Cauchy stress tensor $\mathbb{T}$ necessary to
maintain the finite homogeneous deformation \eqref{x} is given by
\cite{BoHT94}
\begin{equation} \label{T}
\mathbb{T}=  [ a II \, III^{-1/2}+ g(III^{1/2})] \mathbf{1}
 + b III^{-1/2} \mathbb{B} - a III^{1/2} \mathbb{B}^{-1}.
\end{equation}

\section{Small-amplitude plane waves in a deformed Hadamard material}
\label{Small-amplitude plane waves in a deformed Hadamard material}

Now we consider the propagation of an infinitesimal plane wave of
complex exponential type in a Hadamard material held in a state of
static finite homogeneous deformation.
The emphasis is on inhomogeneous plane waves.
We derive the equations of motion and the corresponding acoustical
tensor.

\subsection{Plane waves of complex exponential type}

We assume that a plane wave of complex exponential type is superposed
upon the finite static deformation described in \S \ref{Homogeneous
Deformation}. This motion, which brings a particle from $\mathbf{x}$,
given by \eqref{x}, to $\mathbf{\overline{x}}$ in the current
configuration of the material, is written as  \cite{Haye84}
\begin{equation} \label{xBar}
\mathbf{\overline{x}} = \mathbf{x}+ \textstyle{\frac{1}{2}} \epsilon \{
\mathbf{A} e^{i \omega(N\mathbf{C \cdot x}-t)} +\mathrm{c.c.} \}.
\end{equation}
Here, $\epsilon$ is a small parameter, such that terms of order
$\epsilon^2$ or higher may be neglected in comparison with first order
terms, $\mathbf{A}$ (amplitude) and $ \mathbf{C}$ (propagation) are
complex vectors (or `bivectors' \cite{Haye84}), $\omega$ is the real
frequency, $N$ the complex scalar slowness, and `c.c.' denotes the
complex conjugate.

An ellipse may be associated with a bivector \cite{Haye84}.
Thus, if the bivector
$\mathbf{D}$ has real and imaginary parts $\mathbf{D^+}$ and
$\mathbf{D^-}$, so that $\mathbf{D}=\mathbf{D^+}+i\mathbf{D^-}$, then
the equation of the corresponding ellipse is $\mathbf{r}=\mathbf{D^+}
\cos \theta + \mathbf{D^-} \sin \theta, 0 \le \theta \le 2 \pi$. The
ellipse is a circle if $\mathbf{D \cdot D}=0$, and degenerates to
a line segment if $\mathbf{D \wedge \overline{D}}=
\mathbf{0}$.

In the case where the  amplitude bivector $\mathbf{A}$ is such that
\begin{equation}
\mathbf{A \cdot A}=0,
\end{equation}
the wave described by \eqref{xBar} is circularly-polarized and the
amplitude bivector $\mathbf{A}$ is said to be `isotropic'.
If $\mathbf{A \wedge \overline{A}}= \mathbf{0}$, the wave is
linearly-polarized, and $\mathbf{A}$ has `a real direction'.

In general, the propagation bivector $\mathbf{C}$ may be written as
\cite{BoHa93}
\begin{equation} \label{C}
\mathbf{C}= m \mathbf{\widehat{m}} + i \mathbf{\widehat{n}},
\end{equation}
where $m$ is a real number ($m \ge 1$) and
$\mathbf{\widehat{m}},\mathbf{\widehat{n}}$ are real orthogonal unit
vectors. By suitable choices of $m, \mathbf{\widehat{m}}$ and
$\mathbf{\widehat{n}}$, all possible propagation bivectors
$\mathbf{C}$ are determined \cite{Haye84}.

\subsection{Strain increments}

Let the deformation gradient corresponding to the motion \eqref{xBar}
be $\overline{\mathbb{F}}$. It is given by
\begin{equation} \label{Fbar}
\overline{\mathbb{F}} =
\frac{\partial\overline{\mathbf{x}}}{\partial\mathbf{X}} = [ \mathbf{1}
+ \textstyle{\frac{1}{2}} \epsilon \{ i \omega N \mathbf{A} \otimes
\mathbf{C} e^{i \omega (N \mathbf{C \cdot x}-t)} +\mathrm{c.c.} \} ]
\mathbb{F}.
\end{equation}
The left Cauchy--Green tensor
$\overline{\mathbb{B}}=\overline{\mathbb{F}} \
\overline{\mathbb{F}}^{\mathrm{T}}$ and its inverse
$\overline{\mathbb{B}}\,^{-1}$ are, up to order $\epsilon$, given by
\begin{align} \label{Bbar}
&\overline{\mathbb{B}} & = & \: \mathbb{B} + \textstyle{\frac{1}{2}}
\epsilon \{ i \omega N [ \mathbf{A} \otimes \mathbb{B} \mathbf{C} +
\mathbb{B} \mathbf{C} \otimes \mathbf{A} ] e^{i \omega (N \mathbf{C
\cdot  x}-t)}+ \mathrm{c.c.} \},
\\ 
&\overline{\mathbb{B}}\,^{-1} & = &  \: \mathbb{B}^{-1}
-\textstyle{\frac{1}{2}} \epsilon \{ i \omega N [ \mathbf{C} \otimes
\mathbb{B}^{-1} \mathbf{A}
 + \mathbb{B}^{-1} \mathbf{A} \otimes  \mathbf{C} ]
 e^{i \omega (N \mathbf{C \cdot x}-t)} +\mathrm{c.c.} \}.
\end{align}
The corresponding strain invariants
$\overline{I},\overline{II},\overline{III}$ are given by
\begin{align} \label{IbarIIbarIIIbar}
&\overline{I} &  = &  \: I +
 \epsilon \{ i \omega N (\mathbf{A \cdot} \mathbb{B} \mathbf{C})
e^{i \omega (N \mathbf{C \cdot x}-t)} +\mathrm{c.c.} \}, \\
&\overline{II} &  = &  \: II +  \epsilon  \{ i \omega N
[II (\mathbf{A \cdot C}) - III( \mathbf{A \cdot}
\mathbb{B}^{-1} \mathbf{C})]
 e^{i \omega (N \mathbf{C \cdot x}-t)}+ \mathrm{c.c.} \}, \\
&\overline{III} &  = &  \: III 
	+  \epsilon III \{ i \omega N ( \mathbf{A
\cdot C}) e^{i \omega (N \mathbf{C \cdot x}-t)}+ \mathrm{c.c.} \}.
\end{align}
Finally, the mass density  $\overline{\rho}$ in the current
configuration is
\begin{equation} \label{rhoBar}
\overline{\rho} = III^{-1/2} \rho ( 1 - \textstyle{\frac{1}{2}}
\epsilon \{ i \omega N ( \mathbf{A \cdot C}) e^{i \omega (N \mathbf{C
\cdot x}-t)}+ \mathrm{c.c.} \}).
\end{equation}

\subsection{Equations of motion and acoustical tensor}
\label{Equations of motion and acoustical tensor}

The Cauchy stress $\overline{\mathbb{T}}$ necessary to support the
motion \eqref{xBar} is \cite{BoHT94}
\begin{equation} \label{Tbar}
\overline{\mathbb{T}} = [ a \overline{II} \, \overline{III}\,^{-1/2}+
g(\overline{III}^{1/2})]\mathbf{1} + b \overline{III}\,^{-1/2}
\overline{\mathbb{B}} - a \overline{III}^{1/2}
\overline{\mathbb{B}}\,^{-1}.
\end{equation}

Using \eqref{Bbar}--\eqref{Tbar}, we find that
\begin{equation}
\overline{\mathbb{T}} = \mathbb{T}+ \textstyle{\frac{1}{2}} \epsilon \{
i \omega N \widehat{\mathbb{T}} e^{i \omega (N \mathbf{C \cdot x}-t)}
+\mathrm{c.c.} \},
\end{equation}
where $\widehat{\mathbb{T}}$ is given by
\begin{multline} \label{T^}
\widehat{\mathbb{T}} =  \{ (\mathbf{A \cdot C}) g' + a [(\mathrm{tr} \,
\mathbb{B}^{-1}) (\mathbf{A \cdot C}) - 2 (\mathbf{A \cdot}
\mathbb{B}^{-1}\mathbf{C})] \} III^{1/2}   \mathbf{1} \\ - b (\mathbf{A
\cdot C}) III^{-1/2} \mathbb{B} + b III^{-1/2} [ \mathbf{A} \otimes
\mathbb{B} \mathbf{C} + \mathbb{B} \mathbf{C} \otimes \mathbf{A} ] \\ -
a  (\mathbf{A \cdot C}) III^{1/2} \mathbb{B}^{-1} + a III^{1/2}  [
\mathbf{C} \otimes \mathbb{B}^{-1} \mathbf{A} + \mathbb{B}^{-1}
\mathbf{A} \otimes \mathbf{C} ].
\end{multline}

Now, in the absence of body forces, the equations of motion
\eqref{Motion}, written for the motion \eqref{xBar}, are: $\text{ div }
\overline{\mathbb{T}} = \overline{\rho} \
\ddot{\mathbf{\overline{x}}}$. In our context, they yield
\begin{equation}
- \omega^2 N^2 \widehat{\mathbb{T}} \mathbf{ \cdot C}
 = -\rho \omega^2 III^{-1/2} \mathbf{A}, \quad \text{ or }
\quad \mathbb{Q}(\mathbf{C}) \mathbf{A} = \rho N^{-2} \mathbf{A},
\end{equation}
where the \textit{acoustical tensor} $\mathbb{Q}(\mathbf{C})$ is given
by
\begin{multline} \label{Q}
\mathbb{Q}(\mathbf{C}) = b (\mathbf{C \cdot} \mathbb{B} \mathbf{C})
\mathbf{1} + (III g' + a II) \mathbf{C} \otimes  \mathbf{C} \\ + a III
[(\mathbf{C \cdot C})\mathbb{B}^{-1}  -\mathbf{C} \otimes
\mathbb{B}^{-1} \mathbf{C} - \mathbb{B}^{-1} \mathbf{C} \otimes
\mathbf{C}].
\end{multline}
By inspection of the form of $\mathbb{Q} (\mathbf{C})$, we deduce two
facts about the acoustical tensor.

First, the acoustical tensor is a complex symmetric tensor,
\begin{equation}
\mathbb{Q} = \mathbb{Q}^{\mathrm{T}},
\end{equation}
and therefore, eigenbivectors of $\mathbb{Q} (\mathbf{C})$
corresponding to distinct eigenvalues will be orthogonal to each other
(e.g. \cite{BoHa93}).

Second, the propagation bivector $\mathbf{C}$ is an eigenbivector of
$\mathbb{Q} (\mathbf{C})$ with eigenvalue $\rho N_{\parallel}^{-2}$
(say) given by
\begin{equation} \label{rhoN_C}
\rho N_{\parallel}^{-2}=  b (\mathbf{C \cdot} \mathbb{B} \mathbf{C})+
(III g' + a II) (\mathbf{C \cdot C}) - a III (\mathbf{C \cdot
}\mathbb{B}^{-1} \mathbf{C}).
\end{equation}

We  now seek small-amplitude circularly-polarized inhomogeneous
plane wave solutions to the equations of motion in a deformed Hadamard
material.
These solutions correspond to a double root of the acoustical
tensor, or equivalently \cite{Haye84}, to an isotropic amplitude
eigenbivector $\mathbf{A}$ for $\mathbb{Q} (\mathbf{C})$.
We follow the usual \cite{BoHa96,Dest99b} procedure for inhomogeneous
solutions of complex exponential type, which consists in separating
the cases where the propagation bivector $\mathbf{C}$ is
non-isotropic ($\mathbf{C \cdot C} \ne 0$),
from the cases where it is isotropic ($\mathbf{C \cdot C}=0$).

\section{Circularly-polarized inhomogeneous plane waves with
a non-iso\-tro\-pic propagation bi\-vec\-tor $\mathbf{C}$}
\label{waves with a non-isotropic bivector}

Here we determine all possible circularly-polarized solutions of
complex exponential type with a non-isotropic bivector
$\mathbf{C}$: $\mathbf{C \cdot C} \ne 0$.
It is seen that these waves can be constructed not only as
`transverse waves' -- in the sense that $\mathbf{A \cdot C}=0$,
but also as the superposition of a `transverse wave' and a
`longitudinal wave' -- in the sense that
$\mathbf{A \wedge C}=\mathbf{0}$.
Explicit solutions are presented.

Assuming that the ellipse of $\mathbf{C}$ is not a circle,
so that $\mathbf{C \cdot C} \ne 0$, the acoustical tensor
$\mathbb{Q}(\mathbf{C})$ may be written
\begin{equation} \label{Qnoniso}
\mathbb{Q}(\mathbf{C}) = b (\mathbf{C \cdot} \mathbb{B} \mathbf{C})
\mathbf{1}  + \Big{[}III g' + a II - a III \mathbf{C^* \cdot}
\mathbb{B}^{-1}\mathbf{C^*} \Big{]} \mathbf{C} \otimes \mathbf{C}
 +  a III (\mathbf{C \cdot C}) \mbox{\boldmath $\chi$},
\end{equation}
where
\begin{align} \label{Pi}
& \mathbf{C^*} = \mathbf{C}/(\mathbf{C \cdot C})^{1/2},
    \quad
\mbox{\boldmath $\chi$} =
    \mbox{\boldmath $\Pi$} \mathbb{B}^{-1} \mbox{\boldmath $\Pi$},
\nonumber \\
& \mbox{\boldmath $\Pi$} =
    \mathbf{1} -  \mathbf{C^*} \otimes \mathbf{C^*},
\quad \mbox{\boldmath $\Pi$}^2 =  \mbox{\boldmath $\Pi$}, \quad
     \mbox{\boldmath $\Pi$} \mathbf{C} = \mathbf{0}.
\end{align}
Also $\mbox{\boldmath $\Pi$} \mathbf{A} = \mathbf{A}$ when
$\mathbf{A \cdot C} = 0$.
The tensor $\mbox{\boldmath $\Pi$}$ is called the
`complex projection operator'.
Detailed properties of $\mbox{\boldmath $\Pi$}$ 
are given in the Appendix.
In particular, properties of $\mbox{\boldmath $\chi$}$
are presented there.

We recall that $\mathbb{Q}(\mathbf{C})$ has eigenbivector
$\mathbf{C}$ with eigenvalue $\rho N_{\parallel}^{-2}$
given by \eqref{rhoN_C}:
\begin{equation}
\mathbb{Q}(\mathbf{C}) \mathbf{C} = \rho N_{\parallel}^{-2}\mathbf{C}.
\end{equation}

We wish to determine the other eigenbivectors.
Four cases have to be considered (see Appendix):

Case (i): $\mathbf{n^+ \cdot \mbox{\boldmath $\Pi$} n^+} \ne 0$,
        $\mathbf{n^- \cdot \mbox{\boldmath $\Pi$} n^-} \ne 0$;

Case (iia): $\mathbf{n^+ \cdot \mbox{\boldmath $\Pi$} n^+} = 0$,
        $\mathbf{n^- \cdot \mbox{\boldmath $\Pi$} n^-} \ne 0$;

Case (iib):  $\mathbf{n^+\cdot\mbox{\boldmath $\Pi$} n^+ } \ne 0$,
        $\mathbf{n^- \cdot \mbox{\boldmath $\Pi$} n^-} = 0$,

Case (iii): $\mathbf{n^+\cdot\mbox{\boldmath $\Pi$} n^+ } = 0$,
        $\mathbf{n^- \cdot \mbox{\boldmath $\Pi$} n^-} = 0$.

\subsection{Case(i): Superposition of `transverse' and
`longitudinal' waves}
\label{Superposition of `transverse' and `longitudinal' waves}

Here, $\mathbf{n^+ \cdot \mbox{\boldmath $\Pi$} n^+} \ne 0$,
$\mathbf{n^- \cdot \mbox{\boldmath $\Pi$} n^-} \ne 0$.
It follows that the orthogonal projection of the ellipse of
$\mathbf{C}$ upon either plane of central circular section of the
 $\mathbb{B}^{-1}$-ellipsoid is not a circle.
It will be seen that the acoustical tensor has $\mathbf{C}$ as an
eigenbivector, so that the corresponding wave may be called
`longitudinal', and two further distinct orthogonal eigenbivectors
both orthogonal to $\mathbf{C}$, so that the corresponding waves may
be called `transverse'.
The transverse wave slownesses are distinct, so that for a given
$\mathbf{C}$, if two wave slownesses are to be equal, then this
possibility will only occur if the wave slowness for one of the
`transverse' waves is equal to the wave slowness of the `longitudinal'
wave.
It will be seen that there is an infinite number of possible
choices of $\mathbf{C}$ for which this is so.
Then the corresponding waves are circularly-polarized.
Any possible $\mathbf{C}$ has an ellipse which is similar and
similarly situated to the ellipse in which the plane of $\mathbf{C}$
cuts a certain $\mathbb{M}$-ellipsoid, with the two exceptions when
the plane of $\mathbf{C}$ coincides with a plane of central circular
section of the  $\mathbb{M}$-ellipsoid.

It may be shown (see Appendix) that
$\mbox{\boldmath $\chi$}$ has unit orthogonal
eigenbivectors $\mathbf{A^\pm}$, given by
\begin{equation}
\mathbf{A^\pm} =
  (\mathbf{K^+} \pm \mathbf{K^-})/[2(1+ \mathbf{K^+ \cdot K^-})]^{1/2},
\quad
 \mathbf{K^\pm}=
  \mathbf{\mbox{\boldmath $\Pi$} n^\pm}/(\mathbf{n^\pm \cdot  \mbox{\boldmath $\Pi$} n^\pm})^{1/2},
\end{equation}
and corresponding distinct eigenvalues $\delta^{\pm}$, given by
\begin{equation}
\delta^{\pm} = \lambda_2^{-2}
 - \textstyle{\frac{1}{2}}(\lambda_3^{-2} - \lambda_1^{-2})
 [(\mathbf{n^+ \cdot \mbox{\boldmath $\Pi$} n^+})(\mathbf{n^- \cdot \mbox{\boldmath $\Pi$} n^-})]^{1/2}
 (\mathbf{K^+ \cdot K^-} \pm 1).
\end{equation}
Also, $\mathbf{A}^{\pm}$ are orthogonal to $\mathbf{C}$, the other
eigenbivector of
$\mbox{\boldmath $\chi$}$, with corresponding eigenvalue zero.

So the eigenbivectors of $\mathbb{Q}(\mathbf{C})$ are:
$\mathbf{C}$, with eigenvalue $\rho N_{\parallel}^{-2}$, and
$\mathbf{A}^{\pm}$, with corresponding distinct eigenvalues
 $b \mathbf{C \cdot} \mathbb{B} \mathbf{C}
    + a III (\mathbf{C \cdot C}) \delta^{\pm}$.

Because the eigenvalues corresponding to $\mathbf{A^+}$ and
 $\mathbf{A^-}$ are distinct, the only way in which an isotropic
eigenbivector of $\mathbb{Q}(\mathbf{C})$ will arise is when the
 eigenvalue $\rho N_{\parallel}^{-2}$ is equal to one of the
eigenvalues corresponding to either $\mathbf{A^+}$ or to
$\mathbf{A^-}$.
Thus, for an isotropic eigenbivector we need $\mathbf{C}$ to satisfy
\begin{equation}
[\rho N_{\parallel}^{-2} - b \mathbf{C \cdot} \mathbb{B} \mathbf{C}
    - a III (\mathbf{C \cdot C}) \delta^+]
[\rho N_{\parallel}^{-2} - b \mathbf{C \cdot} \mathbb{B} \mathbf{C}
    - a III (\mathbf{C \cdot C}) \delta^-] = 0.
\end{equation}
Using \eqref{rhoN_C} and \eqref{delta+-properties}, 
this equation may be written
\begin{equation} \label{C.MC}
\mathbf{C \cdot} \mathbb{M} \mathbf{C}=0,
\end{equation}
where $\mathbb{M}$ is the tensor defined by
\begin{equation}
\mathbb{M}= (g')^2 III \mathbf{1} +
 a g' [II \mathbf{1} -III  \mathbb{B}^{-1}]
+ a^2 \mathbb{B}.
\end{equation}
We note that  $\mathbb{M}$ is coaxial with $\mathbb{B}$ and has
eigenvalues $\mu^2_\alpha$, given by
\begin{equation} \label{mu}
\mu^2_\alpha= (g')^2 III +
    a g' ( II - III \lambda^{-2}_\alpha)
        + a^2 \lambda^2_\alpha \quad (\alpha=1,2,3).
\end{equation}
Using \eqref{S-E} and \eqref{orderLambda}, we find that
\begin{equation} \label{orderMu}
\mu_1^2>\mu_2^2>\mu_3^2>0.
\end{equation}
Thus, $\mathbb{M}$ is a positive definite tensor,
determined by the finite static deformation,
and the associated quadric $\mathbf{x \cdot} \mathbb{M} \mathbf{x}=1$
is an ellipsoid.

So now if $\mathbf{C}$ is chosen to satisfy  \eqref{C.MC},
that is, if the plane of the ellipse of $\mathbf{C}$ cuts the
$\mathbb{M}$-ellipsoid in an ellipse which is similar and
similarly situated to the ellipse of $\mathbf{C}$,
then $\mathbb{Q}(\mathbf{C})$ has a double eigenvalue,
$\rho N^{-2}_\parallel$, corresponding to the eigenbivector
$\mathbf{C}$, and also either to $\mathbf{A^+}$ or to
$\mathbf{A^-}$.
Consequently, all isotropic eigenbivectors $\mathbf{A}$ of the
acoustical tensor $\mathbb{Q}(\mathbf{C})$ must be orthogonal to
$\mathbf{A^+}$ (or $\mathbf{A^-}$), be such that
$\mathbf{A \cdot A}=0$, and consequently, up to a scalar factor,
be of the form
\begin{equation} \label{Atrans+long}
\mathbf{A}= \mathbf{C^*} +  i \mathbf{A^-}, \quad
(\text{or} \quad
\mathbf{C^*} +  i \mathbf{A^+}),
\end{equation}
with corresponding eigenvalue $\rho N^{-2}_\parallel$.
Of course, the amplitude bivector $\mathbf{A}$ is a combination
of $\mathbf{C^*}$, which corresponds to a `longitudinal' wave
in the sense that the amplitude bivector is `parallel' to the
propagation bivector $\mathbf{C}$, and of the amplitude
 $\mathbf{A^-}$, which corresponds to a `transverse' wave
in the sense that the amplitude bivector $\mathbf{A^-}$ is
`orthogonal' to the propagation bivector $\mathbf{C}$.

There is an infinity of possible choices for $\mathbf{C}$ which
satisfy   \eqref{C.MC}.
Any elliptical section of the $\mathbb{M}$-ellipsoid may be chosen for
$\mathbf{C}$, apart from the two central circular sections -- we
recall that $\mathbf{C}$ may not be isotropic.
These `forbidden' isotropic $\mathbf{C}$, in the planes of central
circular section of the $\mathbb{M}$-ellipsoid, are
\begin{equation}
\mathbf{C} =
 \frac{\gamma (a+ \lambda_1^2 g')^{1/2}}
    {\lambda_1 (a+ \lambda_2^2 g')^{1/2}} \mathbf{i}
 \pm i\mathbf{j}
 \mp  \frac{\alpha (a+ \lambda_3^2 g')^{1/2}}
    {\lambda_3 (a+ \lambda_1^2 g')^{1/2}} \mathbf{k}.
\end{equation}

\subsection{Example: Superposition of `transverse' and `longitudinal'
waves}

Let $\mathbf{C}$ be written
$\mathbf{C}= C_1 \mathbf{i} + C_2 \mathbf{j} + C_3 \mathbf{k}$.
The condition \eqref{C.MC} is
$C_1^2 \mu_1^2 + C_2^2 \mu_2^2 + C_3^2 \mu_3^2 = 0$.
Recalling \eqref{orderMu} we choose
\begin{equation} \label{Cmu}
\mathbf{C}= \mu_2 \mathbf{i} + i \mu_1 \mathbf{j}.
\end{equation}
Then
\begin{equation}
\mbox{\boldmath $\Pi$} \mathbf{n^\pm} =
  - \frac{\alpha \mu_1^2}{\mu_2^2-\mu_1^2} \mathbf{i}
  \pm \gamma \mathbf{k}
  - i\frac{\alpha \mu_1 \mu_2}{\mu_2^2-\mu_1^2} \mathbf{j}.
\end{equation}
It may be checked that  $\mathbf{n^+ \cdot \mbox{\boldmath $\Pi$} n^+}
 = \mathbf{n^- \cdot \mbox{\boldmath $\Pi$} n^-}$, so that
$\mathbf{A^\pm}$ are parallel to
$\mbox{\boldmath $\Pi$} \mathbf{n^+}
  \pm \mbox{\boldmath $\Pi$} \mathbf{n^-}$.
Thus,
\begin{align}
& \mathbf{A^+}=
  (\mu_1 \mathbf{i} - i\mu_2\mathbf{j})/(\mu_1^2-\mu_2^2)^{1/2},
 \quad
\mathbf{A^-}= \mathbf{k},  \nonumber \\
& \delta^+ = \mathbf{A^+ \cdot}\mbox{\boldmath $\chi$} \mathbf{A^+}
    = \mathbf{A^+ \cdot} \mathbb{B}^{-1} \mathbf{A^+}
 = (\mu_1^2 \lambda_1^{-2} -\mu_2^2 \lambda_2^{-2})/(\mu_1^2-\mu_2^2),
 \\
&\delta^- = \mathbf{A^- \cdot}\mbox{\boldmath $\chi$} \mathbf{A^-}
    = \mathbf{A^- \cdot} \mathbb{B}^{-1} \mathbf{A^-}
 =  \lambda_3^{-2}.\nonumber
\end{align}

Using \eqref{mu}, we have
\begin{align} \label{Nexpl2}
& \mathbf{C \cdot C} = \mu_2^2 - \mu_1^2
    = a(\lambda_2^2 -\lambda_1^2)(a + g' \lambda_3^2),
\nonumber \\
& \mathbf{C \cdot} \mathbb{B}^{-1} \mathbf{C} =
    \mu_2^2\lambda_1^{-2} - \mu_1^2 \lambda_2^{-2} \\
& \phantom{\mathbf{C \cdot C} =} = (\lambda_2^2 -\lambda_1^2)
  [(g')^2 III + ag' II + a^2 (\lambda_1^2
    +\lambda_2^2)]/(\lambda_1^2\lambda_2^2), \nonumber \\
& \rho N_\parallel^{-2} =
 (\lambda_2^2 -\lambda_1^2) III (g' + a\lambda_3^{-2})(a^2 -b g').
\nonumber
\end{align}
It then follows that the difference between the eigenvalues
corresponding to $\mathbf{C}$ and $\mathbf{A^-}$ is
\begin{multline}
 \rho N_\parallel^{-2}
  - b\mathbf{C \cdot} \mathbb{B}^{-1} \mathbf{C}
   - a III \lambda_3^{-2} \mathbf{C \cdot C} \\
= (III g' + a II) \mathbf{C \cdot C}
    - aIII \mathbf{C \cdot} \mathbb{B}^{-1} \mathbf{C}
        - aIII  \lambda_3^{-2} \mathbf{C \cdot C}= 0.
\end{multline}

Thus the isotropic eigenbivector corresponding to the eigenvalue
$\rho N_\parallel^{-2}$ for the above choice \eqref{Cmu} of
$\mathbf{C}$ is
\begin{equation} \label{Aexpl2}
\mathbf{A}=
  (\mu_2\mathbf{i} +i \mu_1\mathbf{j})/(\mu_1^2-\mu_2^2)^{1/2}
  \pm i \mathbf{k}.
\end{equation}
Then, assuming that $a^2 -b g'>0$,  the corresponding explicit
solution is given by
\begin{align} \label{Expl2.1}
& \overline{x} = x+ \epsilon [\mu_2/(\mu_2^2-\mu_1^2)^{1/2}]
 e^{- \omega N_\parallel \mu_1 y}\cos \omega (N_\parallel \mu_2 x- t),
 \nonumber \\
& \overline{y} =y+ \epsilon [\mu_1/(\mu_2^2-\mu_1^2)^{1/2}]
 e^{- \omega N_\parallel \mu_1 y}\cos \omega (N_\parallel \mu_2 x- t),
\\
& \overline{z} = z \pm \epsilon  e^{- \omega N_\parallel \mu_1 y}
\sin \omega (N_\parallel \mu_2 x- t).
\nonumber
\end{align}
Here ($x,y,z$)=($\lambda_1 X, \lambda_2Y, \lambda_3 Z$),
$N_\parallel$ is given by \eqref{Nexpl2}$_3$,
and $\mu_1, \mu_2$ by \eqref{mu}.
One radius of the circle of polarization is along $\mathbf{i}$,
an orthogonal radius along
$\mu_1 \mathbf{j} \pm (\mu_1^2-\mu_2^2)^{1/2} \mathbf{k}$.
The waves travel in the $x$-direction with speed
$(N_\parallel \mu_2)^{-1}$, and are
attenuated in the $y$-direction.

\subsection{Remark. Special case $\mathbf{C \cdot j} =0$}

It is shown  (Appendix, Case (i)) that
$\mbox{\boldmath $\Pi$} \mathbf{n^+}$ and
$\mbox{\boldmath $\Pi$} \mathbf{n^-}$ are parallel if
$\mathbf{C \cdot j} =0$.
The eigenbivectors of $\mbox{\boldmath $\chi$}$ are then
$\mathbf{C^*}$, $\mathbf{K^+}$, $\mathbf{j}$.
Using \eqref{Qnoniso}, the eigenbivectors of $\mathbb{Q}(\mathbf{C})$
are also $\mathbf{C^*}$, $\mathbf{K^+}$, $\mathbf{j}$ in this case.
However, because $C_2^*=0$, $(C_1^*)^2 + (C_3^*)^2 = 1$,
it follows that equating any two of the eigenvalues of
 $\mathbb{Q}(\mathbf{C})$ will lead to a $\mathbf{C^*}$ which is a
`real' bivector -- a scalar multiple of a real vector.

The conclusion is therefore that if $\mathbf{C \cdot j} =0$,
then the only possible circularly-polarized waves are homogeneous.

\subsection{Case(iia): `Transverse' circularly-polarized waves}

Here, $\mathbf{n^- \cdot \mbox{\boldmath $\Pi$} n^-} \ne 0$ and
$\mathbf{n^+ \cdot \mbox{\boldmath $\Pi$} n^+} = 0$, so that
$\mbox{\boldmath $\Pi$} \mathbf{n^+}$ is isotropic.
The projection, of the ellipse of $\mathbf{C}$ upon the plane of
central circular section of the $\mathbb{B}^{-1}$-ellipsoid with
normal $\mathbf{n^+}$, is a circle.

Using the results in the Appendix relating to Case (iia),
we conclude that $\mbox{\boldmath $\Pi$} \mathbf{n^+}$ is an
eigenbivector of  $\mathbb{Q}(\mathbf{C})$ with eigenvalue
$\rho N_{\perp}^{-2}$, given by
\begin{align} \label{rhoN_perp}
& \rho N^{-2}_{\perp}= b (\mathbf{C \cdot} \mathbb{B} \mathbf{C}) +
a III (\mathbf{C \cdot C}) \nu_{\perp},
\nonumber \\
& \nu_{\perp}   = \lambda_2^{-2}
    - \textstyle{\frac{1}{2}}(\lambda_3^{-2}-\lambda_1^{-2})
\mathbf{n^+} \cdot \mbox{\boldmath $\Pi$}\mathbf{n^-}
= \textstyle{\frac{1}{2}}I_{\mbox{\boldmath $\chi$}}.
\end{align}
See \eqref{alphaHat}.
The eigenvalue $\rho N_{\perp}^{-2}$ is a double root of the secular 
equation because one root is zero and the sum of the other roots is
$2 b  (\mathbf{C \cdot} \mathbb{B} \mathbf{C}) +
a III (\mathbf{C \cdot C}) I_{\mbox{\boldmath $\chi$}}
 = 2 \rho N_{\perp}^{-2}$.
All isotropic eigenbivectors, $\mathbf{A_{\perp}}$ (say),
of $\mathbb{Q}(\mathbf{C})$, corresponding to $\rho N^{-2}_{\perp}$,
are parallel to $\mathbf{\mbox{\boldmath $\Pi$} n^+}$,
and orthogonal to $\mathbf{C}$ (see Appendix).
Because $\mathbf{C \cdot A_{\perp}}=0$, we refer to these waves
as `transverse'.

Using the fact that
$\mathbf{n^+ \cdot \mbox{\boldmath $\Pi$} n^+} = 0$, we deduce that
$\mathbf{C}$ has the form
\begin{equation}
\mathbf{C} = p \mathbf{n^+} + s(\mathbf{j \wedge n^+} \pm i\mathbf{j}),
\end{equation}
where $p$ and $s$ are arbitrary constants.
To relate this to the expression \eqref{C} for
$\mathbf{C} = m \mathbf{\widehat{m}} + i \mathbf{\widehat{n}}$,
for which $\mathbf{C \cdot C} = m^2 - 1$,
 $\mathbf{C \cdot \overline{C}} = m^2 + 1$, we conclude that
\begin{align} \label{SolnPerp}
\mathbf{C} & =
  \sqrt{m^2 -1} \mathbf{n^+}
    + e^{i \theta}(\mathbf{j \wedge n^+} \pm i\mathbf{j})
\nonumber \\
& =(\alpha \sqrt{m^2 -1} + \gamma  e^{i \theta}) \mathbf{i}
    \pm i e^{i \theta}\mathbf{j} +
     (\gamma \sqrt{m^2 -1} - \alpha e^{i \theta}) \mathbf{k},
\end{align}
where $\theta$ and $m$ are arbitrary.

Also, because $\mathbf{ A_{\perp}}$ is parallel to
$\mathbf{\mbox{\boldmath $\Pi$} n^+}$, it may be written
\begin{equation} \label{Aperp}
\mathbf{A}_\perp= e^{i\theta}(\mathbf{j \wedge n^+} + i\mathbf{j}),
\end{equation}

Now, using \eqref{n+-}, we find that
\begin{equation}
\mathbf{C \cdot} \mathbb{B} \mathbf{C} = III\Lambda_\perp^2,
\quad
\nu_{\perp} =  \sqrt{m^2 -1}\Lambda_\perp,
\end{equation}
where
\begin{equation}  \label{LambdaPerp}
\Lambda_\perp = \sqrt{m^2 -1} \lambda_2^{-2}
+e^{i \theta}\sqrt{\lambda_2^{-2}-\lambda_1^{-2}}
    \sqrt{\lambda_3^{-2}-\lambda_2^{-2}}.
\end{equation}
Hence, from \eqref{rhoN_perp},
\begin{equation}
\rho N^{-2}_{\perp}=
III \Lambda_\perp (b \Lambda_\perp +a \sqrt{m^2 -1}).
\end{equation}

\vspace{10pt}

\textit{Remark: `Forbidden' choices of $\mathbf{C}$.}

Throughout, it has been assumed that
$\mathbf{n^+ \cdot \mbox{\boldmath $\Pi$} n^+} = 0$ and
$\mathbf{n^- \cdot \mbox{\boldmath $\Pi$} n^-} \ne 0$,
and also $\mathbf{C \cdot C} \ne 0$.
The condition $\mathbf{C \cdot C} \ne 0$ means that $\mathbf{C}$
given by \eqref{SolnPerp} must be such that $m \ne 1$,
or equivalently, $\mathbf{C \cdot n^+} \ne 0$.
The propagation bivector $\mathbf{C}$ given by \eqref{SolnPerp} always
satisfies $\mathbf{n^+ \cdot \mbox{\boldmath $\Pi$} n^+} = 0$.
However, as shown in the Appendix, Case (iii),
$\mathbf{n^- \cdot \mbox{\boldmath $\Pi$} n^-} = 0$ also,
if either  $\mathbf{C \cdot i} = 0$ or  $\mathbf{C \cdot k} = 0$.
There are thus three `forbidden' choices of $\mathbf{C}$ given by
\eqref{SolnPerp}: $\mathbf{C \cdot i}= 0$,
or $\mathbf{C \cdot k} = 0$,
or $\mathbf{C \cdot n^+} = 0$.

Any other choices of $\mathbf{C}$ given by \eqref{SolnPerp} lead to
circularly-polarized `transverse' waves, all of which have the common
amplitude bivector $\mathbf{A}_{\perp}$ given by \eqref{Aperp}.
We note that $\mathbf{A}_\perp$ is determined by  $\mathbf{n^+}$,
which, in turn, is determined by the basic static homogeneous
deformation.
There is thus an infinity of circularly-polarized `transverse'
waves all propagating with common amplitude bivector
$\mathbf{A}_{\perp}$ given by \eqref{Aperp}.

\vspace{10pt}

We now present an example.

\subsection{Example: `Transverse' circularly-polarized waves}

We take $\theta =0$, and then from equations
\eqref{SolnPerp}-\eqref{LambdaPerp},
we deduce the following circularly-polarized plane wave solution,
\begin{align} \label{Expl1}
& \overline{x} = x+ \epsilon \sqrt{
 \frac{\lambda_2^{-2}-\lambda_1^{-2}}{\lambda_3^{-2}-\lambda_1^{-2}}}
 e^{- \omega N_\perp y}
\cos \omega (m N_\perp \mathbf{p \cdot x} - t),
\nonumber \\
& \overline{y} = y - \epsilon e^{- \omega N_\perp y}
 \sin \omega (m N_\perp \mathbf{p \cdot x} - t), \\
& \overline{z} = z - \epsilon \sqrt{
 \frac{\lambda_3^{-2}-\lambda_2^{-2}}{\lambda_3^{-2}-\lambda_1^{-2}}}
  e^{- \omega N_\perp y}
\cos \omega (m N_\perp \mathbf{p \cdot x} - t).
\nonumber
\end{align}
Here ($x,y,z$) are the coordinates in the state of finite static
deformation given by \eqref{x}, $m \ge 1$ is arbitrary,
$N_\perp$ is given by \eqref{SolnPerp}$_3$, with
$\Lambda_\perp =  \lambda_2^{-2}  \sqrt{m^2 -1}
+\sqrt{\lambda_2^{-2}-\lambda_1^{-2}}
   \sqrt{\lambda_3^{-2}-\lambda_2^{-2}}$,
and $\mathbf{p}$ is the unit vector defined by
\begin{equation}
m \mathbf{p} =  \mathbf{j \wedge n^+}
+ \sqrt{m^2 -1} \mathbf{n^+}.
\end{equation}

The radius of the circle of polarization  at ($x,y,z$) is
$\epsilon e^{- \omega N_\perp y}$, two orthogonal radii being along
$\mathbf{j}$ and $\mathbf{j \wedge n^+}$.
This wave propagates with speed $(m N_{\perp})^{-1}$ in the direction
of $\mathbf{p}$, and is attenuated in the direction of $\mathbf{j}$,
orthogonal to $\mathbf{p}$.
Note that \textit{homogeneous} circularly-polarized plane waves can
only travel in the direction $\mathbf{n^\pm}$ of an acoustic axis
\cite{BoHT94}.
In the present example, the normal to the planes of constant phase is
$\mathbf{p}$, which may lie in any direction in the $xz$-plane,
as $m$ varies from $1$ to $\infty$.

\subsection{Case(iiib): Further `transverse' circularly-polarized 
waves}

Here, $\mathbf{n^- \cdot \mbox{\boldmath $\Pi$} n^-} = 0$, $\mathbf{n^+
\cdot \mbox{\boldmath $\Pi$} n^+} \ne 0$.
This is identical to Case (iia) when $\mathbf{n^+}$ and
$\mathbf{n^-}$ are interchanged, that is, when $\gamma$ is replaced
by $-\gamma$.
Accordingly, many details will be omitted.

Using the results in the Appendix relating to Case (iib), we conclude
that the isotropic bivector  $\mbox{\boldmath $\Pi$} \mathbf{n^-}$ is
an eigenbivector of $\mathbb{Q}(\mathbf{C})$ with eigenvalue $\rho
\widehat{N}^{-2}_\perp$ say, given by
\begin{align} \label{rhoN_perp2}
& \rho \widehat{N}^{-2}_{\perp}= b (\mathbf{C \cdot} \mathbb{B}
\mathbf{C}) + a III (\mathbf{C \cdot C})  \widehat{\nu}_{\perp},
\nonumber \\ &  \widehat{\nu}_{\perp}   = \lambda_2^{-2}
    - \textstyle{\frac{1}{2}}(\lambda_3^{-2}-\lambda_1^{-2})
\mathbf{n^+} \cdot \mbox{\boldmath $\Pi$}\mathbf{n^-},
\end{align}
precisely the same form as \eqref{rhoN_perp}, but
different in substance.
All isotropic eigenbivectors $\widehat{\mathbf{A}}_\perp$ (say)
of $\mathbb{Q}(\mathbf{C})$,
corresponding to $\rho  \widehat{N}^{-2}_{\perp}$, are parallel to
$\mathbf{\mbox{\boldmath $\Pi$} n^-}$, and orthogonal to $\mathbf{C}$.

As in Case (iia), we deduce the form of $\mathbf{C}$:
\begin{align}  \label{SolnPerp2}
\mathbf{C} & =  \widehat{p} \mathbf{n^-}
    +  \widehat{s}(\mathbf{j \wedge n^-} \pm i\mathbf{j})
\nonumber \\
 & = (\alpha \sqrt{m^2 -1} - \gamma  e^{i \theta}) \mathbf{i}
    \pm i e^{i \theta}\mathbf{j} -
     (\gamma \sqrt{m^2 -1} + \alpha e^{i \theta}) \mathbf{k}.
\end{align}
Also,
\begin{equation} \label{Aperp2}
 \widehat{\mathbf{A}}_\perp
    = e^{i\theta}(\mathbf{j \wedge n^-} + i\mathbf{j}),
\end{equation}
and
\begin{equation}
\mathbf{C \cdot} \mathbb{B} \mathbf{C} = III \widehat{\Lambda}_\perp^2,
\quad  \widehat{\nu}_{\perp} =  \sqrt{m^2 -1}  \widehat{\Lambda}_\perp,
\end{equation}
where
\begin{equation}  \label{LambdaPerp2}
 \widehat{\Lambda}_\perp = \sqrt{m^2 -1} \lambda_2^{-2} - e^{i
\theta}\sqrt{\lambda_2^{-2}-\lambda_1^{-2}}
    \sqrt{\lambda_3^{-2}-\lambda_2^{-2}}.
\end{equation}
Then
\begin{equation}
\rho  \widehat{N}^{-2}_{\perp}= III  \widehat{\Lambda}_\perp (b
 \widehat{\Lambda}_\perp +a \sqrt{m^2 -1}).
\end{equation}

Thus, as in Case (iib), apart from the three cases when
$\mathbf{C \cdot i} = 0$, or $\mathbf{C \cdot k} = 0$,
or $\mathbf{C \cdot n^-} = 0$,
any choice of $\mathbf{C}$ given by \eqref{SolnPerp2} will lead to a
circularly-polarized transverse wave. There is an infinity of such
waves, all sharing a common amplitude bivector
$\widehat{\mathbf{A}}_{\perp}$ given by \eqref{Aperp2}, which is
determined by the basic static homogeneous deformation.

\subsection{Case(iii): Principal circularly-polarized waves}

Here, $\mathbf{n^+ \cdot \mbox{\boldmath $\Pi$} n^+} = 0$, $\mathbf{n^-
\cdot \mbox{\boldmath $\Pi$} n^-} = 0$, so that both $\mbox{\boldmath
$\Pi$} \mathbf{n^+}$ and  $\mbox{\boldmath $\Pi$} \mathbf{n^-}$ are
isotropic.
As shown in the Appendix, Case (iii), the propagation
bivector $\mathbf{C}$ must satisfy either
\begin{equation}
  \text{(a) } \mathbf{C \cdot i} = 0, \quad  \text{or (b) } \mathbf{C \cdot k} = 0,
\end{equation}
(It is not possible to have both $\mathbf{C \cdot i} = 0$, and
$\mathbf{C \cdot k} = 0$, because then $\mathbf{C} \wedge \mathbf{j} =
\mathbf{0}$, and $( \mathbf{C \cdot n^\pm})^2 =0$,  $\mathbf{C \cdot C}
\ne 0$.)
The possible forms for $\mathbf{C}$ are
\begin{subequations}
\begin{equation} \label{SpecialC(a)}
\mathbf{C}= \mathbf{k} \pm i \alpha \mathbf{j},
 \end{equation}
and
\begin{equation} \label{SpecialC(b)}
\mathbf{C}= \mathbf{i} \pm i \gamma \mathbf{j}.
\end{equation}
\end{subequations}
As shown in the Appendix, $\mathbf{\mbox{\boldmath $\Pi$} n^+}$ and
$\mathbf{\mbox{\boldmath $\Pi$} n^-}$ are linearly independent
eigenbivectors of $\mbox{\boldmath $\chi$}$, with common eigenvalue
$\lambda_2^{-2}$.
So these bivectors $\mathbf{\mbox{\boldmath $\Pi$} n^+}$ and
$\mathbf{\mbox{\boldmath $\Pi$} n^-}$ are isotropic eigenbivectors of
the acoustical tensor $\mathbb{Q}(\mathbf{C})$, with eigenvalues $\rho
N_a^{-2}$, $\rho N_b^{-2}$ (say) given by
\begin{subequations}
\begin{equation} \label{SpecialRho(a)}
  \rho N_a^{-2} = b (\mathbf{C \cdot} \mathbb{B} \mathbf{C}) + a III
(\mathbf{C \cdot C}) \lambda_1^{-2}
 = (b + a \lambda_1^2)\lambda_3^2
    (\lambda_2^2 -\lambda_3^2)/(\lambda_1^2 - \lambda_3^2),
\end{equation}
and
\begin{equation} \label{SpecialRho(b)}
  \rho N_b^{-2} = b (\mathbf{C \cdot} \mathbb{B} \mathbf{C})
    + a III (\mathbf{C\cdot C}) \lambda_3^{-2}
= (b + a\lambda_3^2)\lambda_1^2
    (\lambda_1^2 - \lambda_2^2)/(\lambda_1^2 - \lambda_3^2).
\end{equation}
\end{subequations}

Thus, in Case (a), corresponding to $\mathbf{C}$ given by
\eqref{SpecialC(a)}, there are three eigenbivectors for
$\mathbb{Q}(\mathbf{C})$, namely,  $\mathbf{C}$ given by
\eqref{SpecialC(a)} with eigenvalue zero,
$\mathbf{\mbox{\boldmath $\Pi$} n^+}$ with eigenvalue $\rho N_a^{-2}$,
 and $\mathbf{\mbox{\boldmath $\Pi$} n^-}$,
also with eigenvalue $\rho N_a^{-2}$.
Here, $\mathbf{\mbox{\boldmath $\Pi$} n^+}$ and
$\mathbf{\mbox{\boldmath $\Pi$} n^-}$ are respectively parallel to the
bivectors $\mathbf{A_1}$ and $\mathbf{A_2}$ say, given by
\begin{equation} \label{PrincEigenvec}
  \mathbf{A_1}
     = \gamma \mathbf{i} - \alpha \mathbf{k} \mp i \mathbf{j},
\quad
 \mathbf{A_2}
    = \gamma \mathbf{i} + \alpha \mathbf{k} \pm i \mathbf{j}.
\end{equation}

Similarly, in Case (b), there are three eigenbivectors for
$\mathbb{Q}(\mathbf{C})$ -- $\mathbf{C}$ given by
\eqref{SpecialC(b)}, and $\mathbf{\mbox{\boldmath $\Pi$} n^+}$ and
$\mathbf{\mbox{\boldmath $\Pi$} n^-}$, also with common eigenvalue
$\rho N_b^{-2}$.
Here, the bivectors  $\mathbf{\mbox{\boldmath $\Pi$}
n^+}$ and $\mathbf{\mbox{\boldmath $\Pi$} n^-}$ are again
parallel to the bivectors $\mathbf{A_1}$ and $\mathbf{A_2}$ given by
\eqref{PrincEigenvec}.

It may be noted that there are two circularly-polarized waves with
amplitude bivector along $\mathbf{\mbox{\boldmath $\Pi$} n^+}$ given by
\eqref{PrincEigenvec}, one with propagation bivector given by
\eqref{SpecialC(a)}, and complex slowness given by
\eqref{SpecialRho(a)}, the other with $\mathbf{C}$  given by
\eqref{SpecialC(b)}, and complex slowness given by
\eqref{SpecialRho(b)}.
Similarly there are two circularly-polarized
waves with amplitude  bivector along
$\mathbf{\mbox{\boldmath $\Pi$}n^-}$ given by \eqref{PrincEigenvec}.

Using \eqref{S-E}, we note that $\rho N_a^{-2}$, $\rho N_b^{-2}$ are
both real and positive.
Consequently, the direction of propagation, which in general
is parallel to the real part of $N \mathbf{C}$, see \eqref{xBar},
is here parallel to $\mathbf{C^+}$, that is, $\mathbf{k}$
in Case (a), or $\mathbf{i}$ in Case (b).
Similarly, the direction of attenuation, which in general
is parallel to the imaginary part of $N \mathbf{C}$, see \eqref{xBar},
is here parallel to $\mathbf{C^-}$, that is, $\mathbf{j}$.
Hence, the direction of propagation is either along the principal axis
corresponding to the largest strain or along the principal axis
corresponding to the least strain,
whilst the direction of attenuation is along the intermediate axis.
Such inhomogeneous waves, for which the planes of constant phase and
the planes of constant amplitude are normal to principal axes of the
basic homogeneous strain, may be called `principal' waves \cite{Dest00}.
Here, their circle of polarization has a radius along the intermediate axis
and orthogonal radii along either $\mathbf{n^+} \wedge \mathbf{j}$ or
$\mathbf{n^-} \wedge \mathbf{j}$. 
Here is a specific example.

\vspace{10pt}

\textbf{Example: `Principal' circularly-polarized waves}

Let $\mathbf{C}= \mathbf{i} +i \gamma \mathbf{j}$.

In this case, we find the following two special principal circularly
polarized waves,
\begin{align} \label{Expl2}
& \overline{x} = x+ \epsilon \gamma e^{- \omega N_b \gamma y}
\cos \omega (N_b x - t),
\nonumber \\
& \overline{y} = y \pm \epsilon e^{- \omega N_b \gamma y}
 \sin \omega (N_b x - t),
 \\
&\overline{z} = z \mp \epsilon \alpha e^{- \omega N_b \gamma y}
\cos \omega (N_b x - t),
\nonumber
\end{align}
where ($x,y,z$)=($\lambda_1 X, \lambda_2Y, \lambda_3 Z$) and $\omega$
is arbitrary.
For these waves, the propagation is in the direction of $\mathbf{i}$,
the attenuation in the direction of $\mathbf{j}$,
and the speed is $ N_b^{-1}$, where $N_b$ is given by
\eqref{SpecialRho(b)}.

\vspace{10pt}

Hence, we have investigated all possible circularly-polarized
inhomogeneous plane waves of small amplitude propagating in a finitely
deformed Hadamard material, with a non-isotropic bivector $\mathbf{C}$.
There are two types of such solutions.
One type corresponds to waves
whose amplitude bivector $\mathbf{A}$ is orthogonal to the propagation
bivector $\mathbf{C}$; another type corresponds to waves whose
amplitude bivector $\mathbf{A}$ is the sum of a bivector orthogonal to
$\mathbf{C}$ and a bivector parallel to  $\mathbf{C}$.

Central to this investigation was the use of the tensor
$\mbox{\boldmath $\Pi$}$ given by \eqref{Pi}, for which it was
assumed that
$\mathbf{C \cdot C} \ne 0$.
Now we examine in detail the cases where $\mathbf{C \cdot C} = 0$.

\section{Circularly-polarized inhomogeneous plane waves with
an iso\-tro\-pic propagation bivector $\mathbf{C}$} 
\label{waves with an isotropic bivector}

Here we seek circularly-polarized inhomogeneous plane waves
with an iso\-tro\-pic bivector $\mathbf{C}$.
It is seen that such waves can propagate in the deformed Hadamard
material, as long as the circle of $\mathbf{C}$ is not similarly
situated to either of the central circular sections of the
$\mathbb{B}^{-1}$-ellipsoid.

\subsection{The acoustical tensor}
\label{The acoustical tensor(2)}

When $\mathbf{C \cdot C} = 0$, the acoustical tensor
$\mathbb{Q}(\mathbf{C})$ given by \eqref{Q} reduces to
\begin{multline} \label{Qiso}
\mathbb{Q}(\mathbf{C}) = b (\mathbf{C \cdot} \mathbb{B} \mathbf{C})
\mathbf{1} + (III g' + a II) \mathbf{C} \otimes  \mathbf{C} \\
- a III [\mathbf{C} \otimes \mathbb{B}^{-1} \mathbf{C}
           +\mathbb{B}^{-1} \mathbf{C} \otimes \mathbf{C}],
\end{multline}
and $\mathbf{C}$ is an isotropic eigenbivector of
$\mathbb{Q}(\mathbf{C})$ with eigenvalue $\rho N_0^{-2}$
(say), given by
\begin{equation} \label{rhoN_0}
\rho N_0^{-2}=  b (\mathbf{C \cdot} \mathbb{B} \mathbf{C})
 - a III (\mathbf{C \cdot }\mathbb{B}^{-1} \mathbf{C}).
\end{equation}

Because the eigenbivector $\mathbf{C}$ is isotropic, the eigenvalue
 $\rho N_0^{-2}$ is at least double  \cite{BoHa93}.
Let  $\rho N_1^{-2}$ be the remaining eigenvalue of
$\mathbb{Q}(\mathbf{C})$, possibly equal to  $\rho N_0^{-2}$. This
quantity can be deduced from the equality $\text{ tr
}\mathbb{Q}(\mathbf{C})=2 \rho N_0^{-2}+ \rho N_1^{-2}$, using
\eqref{Qiso} and \eqref{rhoN_0}. It is given by
\begin{equation} \label{rhoN_1}
\rho N_1^{-2}=  b (\mathbf{C \cdot} \mathbb{B} \mathbf{C}).
\end{equation}
Also, the bivector $\mathbf{C \wedge} \mathbb{B}^{-1} \mathbf{C}$
 is clearly an eigenbivector of $\mathbb{Q}(\mathbf{C})$,
with eigenvalue $\rho N_1^{-2}$.

Hence, we see that the two eigenvalues given by \eqref{rhoN_0} and
\eqref{rhoN_1} are distinct or equal
according as to whether or not
$\mathbf{C \cdot }\mathbb{B}^{-1} \mathbf{C}$ is equal to zero.
Consequently, we consider in turn the following cases,

(i) $\mathbf{C \cdot }\mathbb{B}^{-1} \mathbf{C} = 0$, $\mathbf{C \cdot
C} = 0$;

(ii) $\mathbf{C \cdot }\mathbb{B}^{-1} \mathbf{C} \ne 0$,  $\mathbf{C
\cdot C} = 0$.

\subsection{Case(i): $ \mathbf{C \cdot C}
=\mathbf{C \cdot }\mathbb{B}^{-1} \mathbf{C} = 0$. \\
`Longitudinal'
circularly-polarized waves}

In this case, $\rho N_0^{-2}=  b (\mathbf{C \cdot} \mathbb{B}
\mathbf{C})$ is a triple eigenvalue of $ \mathbb{Q}( \mathbf{C})$. 
The condition
\begin{equation} \label{C.BC=0}
\mathbf{C \cdot C} = \mathbf{C \cdot }\mathbb{B}^{-1} \mathbf{C} = 0,
\end{equation}
 means that the plane of $\mathbf{C}$ must coincide with either
plane of central circular section of the $\mathbb{B}^{-1}$-ellipsoid 
\cite{BoHa93}. 
Let $\mathbf{A}$ be an eigenbivector of $\mathbb{Q}( \mathbf{C})$. 
Then $\mathbb{Q}( \mathbf{C}) \mathbf{A}= \rho N_0^{-2} \mathbf{A}$ is
equivalent to
\begin{equation} \label{3xEigenval}
 [ a III ( \mathbf{A \cdot C})]
\mathbb{B}^{-1} \mathbf{C} = [(III g' + a II) ( \mathbf{A \cdot C}) - a
III (\mathbf{A \cdot} \mathbb{B}^{-1} \mathbf{C})] \mathbf{C}.
\end{equation}
However, $\mathbb{B}^{-1} \mathbf{C}$ and $ \mathbf{C}$ can never be
parallel. Indeed, if we had $\mathbb{B}^{-1} \mathbf{C}= \lambda^{-2}
\mathbf{C}$, for $ \mathbf{C}= \mathbf{\widehat{m}}+i
\mathbf{\widehat{n}}$, then
$\mathbb{B}^{-1}\mathbf{\widehat{m}}=\lambda^{-2}\mathbf{\widehat{m}}$,
$\mathbb{B}^{-1}\mathbf{\widehat{n}}=\lambda^{-2}\mathbf{\widehat{n}}$,
and $\lambda^{-2}$ would have to be a double real eigenvalue for $
\mathbb{B}^{-1}$, which is not possible. So, the coefficients of
$\mathbb{B}^{-1} \mathbf{C}$ and $ \mathbf{C}$ in \eqref{3xEigenval}
must be zero, which yield in turn
\begin{equation}
\mathbf{A \cdot C}=0, \quad \text{ and } \quad \mathbf{A \cdot}
\mathbb{B}^{-1} \mathbf{C}=0.
\end{equation}
Therefore, $ \mathbf{A}$ is parallel to $  \mathbb{B}^{-1} \mathbf{C
\wedge C}$, which itself is parallel to $ \mathbf{C}$.

We conclude that when \eqref{C.BC=0} holds, all eigenbivectors of $
\mathbb{Q}( \mathbf{C})$ are iso\-tro\-pic and parallel to $ \mathbf{C}$.
The corresponding circularly-polarized waves are said to be
`longitudinal'.

\vspace{10pt}

\textbf{Example: `Longitudinal' circularly-polarized waves}

We let $\mathbf{C}= \mathbf{p} + i \mathbf{j}$, where $\mathbf{p}=
\gamma \mathbf{i} + \alpha \mathbf{k}$.

With this choice, \eqref{C.BC=0} is satisfied, and all amplitude
bivectors are parallel to $\mathbf{C}$. The corresponding eigenvalue
$\rho N_0^{-2}$ is given by
\begin{equation} \label{Nlong}
\rho N_0^{-2} = b (\mathbf{C \cdot} \mathbb{B} \mathbf{C})=
b \lambda_2^{-2}(\lambda_1^2 - \lambda_2^2)(\lambda_2^2 - \lambda_3^2)
 >0.
\end{equation}
Thus an example of a `longitudinal' circularly-polarized wave
propagating in a deformed Hadamard material is
\begin{align} \label{Expl4}
 & \overline{x} = x+ \epsilon \gamma e^{- \omega N_0 y} \cos \omega (N_0
\mathbf{p \cdot x} - t), \nonumber \\
 & \overline{y} = y - \epsilon e^{- \omega N_0
y} \sin \omega (N_0 \mathbf{p \cdot x} - t), \\
 & \overline{z} = z + \epsilon \alpha  e^{- \omega N_0 y}
\cos \omega (N_0 \mathbf{p \cdot x} - t). \nonumber 		
\end{align}
Here ($x,y,z$)=($\lambda_1 X, \lambda_2Y, \lambda_3 Z$) and $\omega$
is arbitrary.
This wave travels in the direction of $\mathbf{p}$ with speed
$N_0^{-1}$ given by \eqref{Nlong}, and is attenuated in the direction
of $\mathbf{j}$.
The vectors $\mathbf{p}$ and $\mathbf{j}$ are two orthogonal
radii of the circle of polarization.

\subsection{Case(ii): $ \mathbf{C \cdot C}= 0$,
$\mathbf{C \cdot }\mathbb{B}^{-1} \mathbf{C} \ne 0$. \\
Other circularly-polarized waves}

Because the ellipse of $\mathbf{C}$ is a circle, the condition
\begin{equation}
\mathbf{C \cdot }\mathbb{B}^{-1} \mathbf{C} \ne 0,
\end{equation}
means that $\mathbf{C}$ may not lie in either plane of central circular
sections of the $\mathbb{B}^{-1}$-ellipsoid. An immediate consequence
of this condition and of the isotropy of $\mathbf{C}$ is that the
bivectors $\mathbf{C}$, $\mathbb{B}^{-1} \mathbf{C}$, and $\mathbf{C
\wedge} \mathbb{B}^{-1} \mathbf{C}$ are linearly independent, so that
any bivector may be written as a linear combination of $\mathbf{C}$,
$\mathbb{B}^{-1} \mathbf{C}$, and $\mathbf{C \wedge} \mathbb{B}^{-1}
\mathbf{C}$. In particular, the eigenbivectors $\mathbf{A}$ of
$\mathbb{Q}(\mathbf{C})$ with eigenvalue $\rho N_0^{-2}$ must be
orthogonal to $\mathbf{C \wedge} \mathbb{B}^{-1} \mathbf{C}$, the
eigenbivector of $ \mathbb{Q}( \mathbf{C})$ with eigenvalue $\rho
N^{-2}_1$ ($\ne \rho N^{-2}_0$), and may therefore be written as
\begin{equation} \label{A_alpha1_alpha2}
\mathbf{A} = \alpha_1 \mathbf{C} + \alpha_2 \mathbb{B}^{-1}\mathbf{C},
\end{equation}
for some complex scalars $\alpha_1$ and $\alpha_2$. We seek to
determine $ \alpha_1$ and $\alpha_2$.

Now, $\mathbb{Q}(\mathbf{C})\mathbf{A} =\rho N_0^{-2}\mathbf{A}$ yields
\begin{equation}
\{ (g') III (\mathbf{C \cdot} \mathbb{B}^{-1} \mathbf{C})
+ a[ II (\mathbf{C \cdot} \mathbb{B}^{-1} \mathbf{C})
    - III (\mathbf{C \cdot} \mathbb{B}^{-2} \mathbf{C})] \}
\alpha_2 =0,
\end{equation}
or, using the Cayley--Hamilton theorem and the isotropy of
$\mathbf{C}$,
\begin{equation}
( \mathbf{C \cdot} \mbox{\boldmath $\Phi$} \mathbf{C}) \alpha_2 =0, 
\text{ where } 
  \mbox{\boldmath $\Phi$}=
 (g') III  \mathbb{B}^{-1} - a \mathbb{B}.
\end{equation}
Therefore, there are two possibilities for the eigenbivectors of
$\mathbb{Q}(\mathbf{C})$:

Case \textbf{(a)} $\alpha_2=0$, and the eigenbivectors
$\mathbf{A}= \alpha_1 \mathbf{C}$ are all isotropic and parallel to
$\mathbf{C}$, and

Case \textbf{(b)}  $\alpha_2 \ne 0$, 
$\mathbf{C \cdot} \mbox{\boldmath $\Phi$} \mathbf{C} =0$, and
there is a double infinity of eigenbivectors, of the form
\eqref{A_alpha1_alpha2}.

Now we examine the consequences of this result for the possibility of
circular polarization.

In Case \textbf{(a)}, all eigenbivectors of the acoustical tensor with
the eigenvalue $\rho N_0^{-2}$ are isotropic and the corresponding
waves are therefore circularly-po\-la\-ri\-zed. 
They are `longitudinal' waves, 
in the sense that the amplitude bivector $\mathbf{A}$ is
parallel to the propagation bivector $\mathbf{C}$.

In Case \textbf{(b)}, $\alpha_2$ is arbitrary and we choose it to make
$ \mathbf{A}$ given by \eqref{A_alpha1_alpha2} isotropic. We have, up
to a complex factor,
\begin{equation}
\mathbf{A}= \mathbf{C} \quad \text{and} \quad
\mathbf{A}=(\mathbf{C \cdot} \mathbb{B}^{-2} \mathbf{C}) \mathbf{C}
- 2 (\mathbf{C \cdot} \mathbb{B}^{-1} \mathbf{C})
              \mathbb{B}^{-1} \mathbf{C}.
\end{equation}
Also, we note that it is always possible to find an isotropic bivector
$\mathbf{C}$ such that the equation 
$\mathbf{C \cdot} \mbox{\boldmath $\Phi$} \mathbf{C} =0$ is
satisfied.
Indeed, because $\mathbf{C \cdot C}=0$, this equation can be
written as $ \mathbf{C \cdot}[ \mbox{\boldmath $\Phi$} + \beta^2
\mathbf{1}]\mathbf{C} =0$, where $\beta$ is an arbitrary real scalar.
By choosing $ \beta$ sufficiently large, we can ensure that the
diagonal tensor $ [\mbox{\boldmath $\Phi$}+ \beta^2 \mathbf{1}]$ is positive
definite, and prescribe the circle of $\mathbf{C}$ to lie in either
of the planes of central circular sections of the ellipsoid
$\mathbf{x\cdot}[\mbox{\boldmath $\Phi$}+ \beta^2 \mathbf{1}] \mathbf{x}=1$.

As an example, we prescribe an isotropic bivector $\mathbf{C}$ and
write the corresponding `longitudinal' circularly-polarized wave.

\vspace{10pt}

\textbf{Example: `Longitudinal' circularly-polarized wave}

Let $\mathbf{C}= \mathbf{q} + i \mathbf{j}$, where the real unit
vector $\mathbf{q}$ is defined by

$\mathbf{q}= \sqrt{
 \frac{(\lambda_2^2-\lambda_3^2)(g' \lambda_1^2 + a)}
        {(\lambda_1^2-\lambda_3^2)(g' \lambda_2^2 + a)}}
\mathbf{i} + \sqrt{
  \frac{(\lambda_1^2-\lambda_2^2)(g' \lambda_3^2 + a)}
        {(\lambda_1^2-\lambda_3^2)(g' \lambda_2^2 + a)}}
\mathbf{k}$.

It can be checked that this bivector $\mathbf{C}$ satisfies
$\mathbf{C \cdot }\mathbb{B}^{-1} \mathbf{C} \ne 0$ and
$ \mathbf{C \cdot C} = \mathbf{C \cdot }\mbox{\boldmath $\Phi$}
  \mathbf{C} = 0$.

The eigenvalue $\rho N_0^{-2}$ given by \eqref{rhoN_0} reduces to
\begin{equation}
\rho N_0^{-2} =
(\lambda_1^2 - \lambda_2^2)(\lambda_2^2 - \lambda_3^2)
 (bg' -a^2)/(g' \lambda_2^2 +a),
\end{equation}
and turns out to be real.
Of course, its sign depends on whether $b g'$ is greater or smaller
than $a^2$.
We consider these two possibilities in turn and introduce the quantity
$v_0$, which has the dimension of a speed, and is defined by
\begin{equation} \label{speedVo}
\rho v_0^2 =
(\lambda_1^2 - \lambda_2^2)(\lambda_2^2 - \lambda_3^2)
 |bg' -a^2|/(g' \lambda_2^2 +a).
\end{equation}

If $b g' > a^2$, then a solution to the incremental equations of
motion in a deformed Hadamard material, corresponding to a
`longitudinal' circularly-polarized wave, is given by
\begin{align} \label{Expl5}
& \overline{x} = x+ \epsilon \sqrt{
 \frac{(\lambda_2^2-\lambda_3^2)(g' \lambda_1^2 + a)}
        {(\lambda_1^2-\lambda_3^2)(g' \lambda_2^2 + a)}}
 e^{- k y}
\cos k ( \mathbf{q \cdot x} - v_0 t),  \nonumber \\
 & \overline{y} = y - \epsilon e^{- k y}
 \sin k ( \mathbf{q \cdot x} - v_0 t), \\
& \overline{z} = z + \epsilon \sqrt{
  \frac{(\lambda_1^2-\lambda_2^2)(g' \lambda_3^2 + a)}
        {(\lambda_1^2-\lambda_3^2)(g' \lambda_2^2 + a)}}
  e^{- k y}
\cos k ( \mathbf{q \cdot x} - v_0  t). \nonumber
\end{align}
If $b g' < a^2$, then a solution is given by
\begin{align} \label{Expl6}
 & \overline{x} = x+ \epsilon \sqrt{
 \frac{(\lambda_2^2-\lambda_3^2)(g' \lambda_1^2 + a)}
        {(\lambda_1^2-\lambda_3^2)(g' \lambda_2^2 + a)}}
 e^{- k \mathbf{q \cdot x} }
\cos k (y - v_0 t),  \nonumber \\
 & \overline{y} = y - \epsilon e^{- k  \mathbf{q\cdot x}}
 \sin k ( y - v_0 t), \\
 & \overline{z} = z + \epsilon \sqrt{
  \frac{(\lambda_1^2-\lambda_2^2)(g' \lambda_3^2 + a)}
        {(\lambda_1^2-\lambda_3^2)(g' \lambda_2^2 + a)}}
  e^{- k \mathbf{q \cdot x} }
\cos k ( y - v_0  t)  \nonumber.
\end{align}
In both cases, ($x,y,z$)=($\lambda_1 X, \lambda_2Y, \lambda_3 Z$)
and $k$ is arbitrary.
When $b g' > a^2$, the wave propagates in the direction of
$\mathbf{q}$ and is attenuated in the direction of $\mathbf{j}$;
when $b g' < a^2$, the wave propagates in the direction of
$\mathbf{j}$ and is attenuated in the direction of $\mathbf{q}$.
In both cases, the wave travels with speed $v_0$ given by
\eqref{speedVo} and is circularly-polarized,
the vectors $\mathbf{q}$ and $\mathbf{j}$ being two orthogonal
radii of the circle of polarization.

\section{Concluding remarks: homogeneous plane waves}
\label{Circularly-polarized homogeneous plane waves}

The analysis carried above
(\S \ref{waves with a non-isotropic bivector} and
\S \ref{waves with an isotropic bivector}) can be applied to the
consideration of homogeneous plane waves, simply by taking
\begin{equation}
\mathbf{C}=\mathbf{n},
\end{equation}
where $\mathbf{n}$ is a real unit vector in the direction of
propagation of the wave.

However, the case of an isotropic propagation vector
$\mathbf{C \cdot C} = 0$ (\S \ref{waves with an isotropic bivector})
does not arise, because now
$\mathbf{C \cdot C} = \mathbf{n \cdot n} = 1$

Also, transverse and longitudinal waves cannot be superposed 
(\S \ref{Superposition of `transverse' and `longitudinal' waves})
to form a circularly-polarized homogeneous wave
because the condition \eqref{C.MC}, which reduces here to
$\mathbf{C \cdot} \mathbb{M} \mathbf{C}
=\mathbf{n \cdot} \mathbb{M} \mathbf{n}
= n_1^2 \mu_1^2 +n_2^2 \mu_2^2 + n_3^2 \mu_3^2   =0$,
cannot be satisfied.

Now, for completeness, we consider briefly the propagation of 
circularly-polarized homogeneous plane waves.

First we recall that homogeneous longitudinal plane waves may 
propagate in every direction $\mathbf{n}$ in a Hadamard material 
maintained in a state of finite static homogeneous deformation.
Two transverse plane waves may also propagate in every direction
$\mathbf{n}$.
If the corresponding directions of polarization are along
$\mathbf{h}, \mathbf{l}$, forming an orthonormal triad with
$\mathbf{n}$, then it has been shown \cite{BoHT94} that
$\mathbf{h \cdot} \mathbb{B}^{-1} \mathbf{l}=0$, so that
$\mathbf{h}$ and $\mathbf{l}$ must lie along the principal axes of the
elliptical section of the $\mathbb{B}^{-1}$-ellipsoid
$\mathbf{x \cdot} \mathbb{B}^{-1} \mathbf{x}=1$ by the central plane
$\mathbf{n \cdot x}=0$.
Here we present a simple derivation of this result.

The acoustical tensor $\mathbb{Q}(\mathbf{n})$ 
for homogeneous plane waves may be obtained from the expression 
\eqref{Q} for $\mathbb{Q}(\mathbf{C})$, by replacing $\mathbf{C}$ 
with $\mathbf{n}$.
Indeed
\begin{multline} \label{Qhom}
\mathbb{Q}(\mathbf{n})
 = b (\mathbf{n \cdot} \mathbb{B} \mathbf{n})\mathbf{1} 
 + (III g' + a II)  \mathbf{n} \otimes \mathbf{n} \\
 +  a III (\mathbb{B}^{-1} 
	- \mathbf{n}\otimes \mathbb{B}^{-1}\mathbf{n}
	- \mathbb{B}^{-1}\mathbf{n}\otimes \mathbf{n}),
\end{multline}
and
\begin{equation} 
\mathbb{Q}(\mathbf{n})\mathbf{n}
 = [b (\mathbf{n \cdot} \mathbb{B} \mathbf{n}) 
 + III g' + a II - a III (\mathbf{n \cdot} \mathbb{B}^{-1}\mathbf{n})]
 \mathbf{n}.
\end{equation}

Because $\mathbf{h}$, $\mathbf{l}$ are eigenvectors of
$\mathbb{Q}(\mathbf{n})$, both orthogonal to $\mathbf{n}$, we have
\begin{align} 
&\mathbb{Q}(\mathbf{n})\mathbf{h}
 = [b (\mathbf{n \cdot} \mathbb{B} \mathbf{n}) 
 +  a III (\mathbf{h \cdot} \mathbb{B}^{-1}\mathbf{h})]
 \mathbf{h}, 
\\
&\mathbb{Q}(\mathbf{n})\mathbf{l}
 = [b (\mathbf{n \cdot} \mathbb{B} \mathbf{n}) 
 +  a III (\mathbf{l \cdot} \mathbb{B}^{-1}\mathbf{l})]
 \mathbf{l}.
\end{align}
Thus, for circularly-polarized homogeneous plane waves, $\mathbf{n}$
must be such that
\begin{equation} 
\mathbf{h \cdot} \mathbb{B}^{-1}\mathbf{h}
	= \mathbf{l \cdot} \mathbb{B}^{-1}\mathbf{l},
\quad
\mathbf{h \cdot} \mathbb{B}^{-1}\mathbf{l} = 0, 
\quad
\mathbf{h \cdot l} = 0,
\end{equation}
and hence, $\mathbf{n} = \mathbf{n^\pm}$, the normals to the planes
of central circular sections of the $\mathbb{B}^{-1}$-ellipsoid.
The corresponding speed of propagation $v_\perp$ say, 
is given by 
\begin{equation} \label{speedHom}
\rho v_\perp^2 = b (\mathbf{n \cdot} \mathbb{B} \mathbf{n})
+ a III \lambda_2^{-2}=
\lambda_2^{-2}(b \lambda_2^{-2} + a III).
\end{equation}
This result was established for finite-amplitude plane
waves by Boulanger \textit{et al} \cite{BoHT94}.
We note that on replacing $\mathbf{C}$ by $\mathbf{n^\pm}$ in 
\eqref{rhoN_perp2}, $\nu_\perp$ becomes $\lambda_2^{-2}$, 
and then we obtain \eqref{speedHom} on replacing $N_\perp^{-2}$
by $v_\perp^2$.



\appendix
 \renewcommand{\thesection}{\Alph{section}}
\small{
\section*{Appendix: Properties of the complex `projection operator'.}

\stepcounter{section}

Here we present some properties of the complex projection operator
$\mbox{\boldmath $\Pi$}$, defined in \eqref{Pi}.
In particular, we determine the eigenvalues and eigenbivectors of
$\mbox{\boldmath $\chi$}
 = \mbox{\boldmath $\Pi$}\mathbb{B}^{-1} \mbox{\boldmath $\Pi$}$,
the tensor which arises in the expression \eqref{Qnoniso} for the
acoustical tensor $\mathbb{Q}(\mathbf{C})$.
Many of these properties may be found in \cite{Dest99b},
but are included here for completeness.
The material is self-contained.

If $ \mathbf{h}$ is a real unit vector, then $\mbox{\boldmath $\phi$}$
defined by
\begin{equation}\label{phi}
  \mbox{\boldmath $\phi$}
    = \mathbf{1} - \mathbf{h} \otimes \mathbf{h}, \quad
  \phi_{ij} = \delta_{ij} - h_i h_j,
\end{equation}
is a projection operator with the properties
\begin{equation}
  \mbox{\boldmath $\phi$}^2 = \mbox{\boldmath $\phi$}, \quad
    \mbox{\boldmath $\phi$}\mathbf{h}= 0;
    \quad \mbox{\boldmath $\phi$} \mathbf{l}= \mathbf{l},
    \quad   \forall \: \mathbf{l}: \mathbf{l \cdot h} = 0.
\end{equation}
For any second order tensor $ \mathbf{g}$, the projection of
$\mathbf{g}$ on the plane with unit normal $\mathbf{h}$, is
$\mbox{\boldmath $\phi$}\mathbf{g} \mbox{\boldmath $\phi$}$,
with components
\begin{equation}
(\mbox{\boldmath $\phi$} \mathbf{g}  \mbox{\boldmath $\phi$})_{ij} =
\phi_{im} g_{mp} \phi_{pj}
    = (\delta_{im} - h_i h_m)g_{mp}(\delta_{pj}- h_p h_j).
\end{equation}

In the case of bivectors, let $ \mathbf{C}$ be a non-isotropic
bivector: $ \mathbf{C \cdot C} \ne 0$.
Let $ \mathbf{C^*}= \mathbf{C}/(\mathbf{C \cdot C})^{1/2}$.
Then the complex `projection operator' $\mbox{\boldmath $\Pi$}$,
defined by
\begin{equation}
  \mbox{\boldmath $\Pi$}
    = \mathbf{1} - \mathbf{C^*} \otimes \mathbf{C^*},
\end{equation}
is such that corresponding to \eqref{phi} for the real projection
operator $\mbox{\boldmath $\phi$}$, we have
\begin{equation}
  \mbox{\boldmath $\Pi$}^2 = \mbox{\boldmath $\Pi$}, \quad
    \mbox{\boldmath $\Pi$}\mathbf{C^*}= 0;
    \quad \mbox{\boldmath $\Pi$} \mathbf{D}= \mathbf{D},
     \forall \mathbf{D}: \mathbf{D \cdot C} = 0.
\end{equation}

We consider the complex tensor $\mbox{\boldmath $\chi$}$, defined by
\begin{equation} \label{chi2}
  \mbox{\boldmath $\chi$} = \mbox{\boldmath $\Pi$}
\mathbb{B}^{-1} \mbox{\boldmath $\Pi$}, \quad \chi_{ij} =  (\delta_{im}
- C^*_i C^*_m)\mathbb{B}^{-1}_{mp}(\delta_{pj} - C^*_p C^*_j),
\end{equation}
where $ \mathbb{B}^{-1}$ is real, positive definite, and symmetric.

\vspace{10pt}
 \textbf{General Properties of $\mbox{\boldmath $\chi$}$:}

Recalling equation \eqref{Hamilton}
\begin{equation} \label{Hamil2}
\mathbb{B}^{-1} = \lambda_2^{-2} \mathbf{1} - \textstyle{\frac{1}{2}}
(\lambda_3^{-2} - \lambda_1^{-2})
    [\mathbf{n}^+ \otimes \mathbf{n}^- +
         \mathbf{n}^- \otimes \mathbf{n}^+ ],
\end{equation}
it follows that
\begin{equation} \label{chi3}
\mbox{\boldmath $\chi$} = \lambda_2^{-2} \mbox{\boldmath $\Pi$} -
\textstyle{\frac{1}{2}} (\lambda_3^{-2} - \lambda_1^{-2})
    [\mbox{\boldmath $\Pi$}\mathbf{n}^+ \otimes \mbox{\boldmath
$\Pi$}\mathbf{n}^- + \mbox{\boldmath $\Pi$}\mathbf{n}^- \otimes
\mbox{\boldmath $\Pi$}\mathbf{n}^+ ].
 \end{equation}
Equivalently, provided $ \mathbf{n^+ \cdot} \mbox{\boldmath
$\Pi$}\mathbf{n^+} \ne 0$, $ \mathbf{n^- \cdot} \mbox{\boldmath
$\Pi$}\mathbf{n^-} \ne 0$, $ \mbox{\boldmath $\chi$}$ may be written
\begin{equation} \label{chi4}
\mbox{\boldmath $\chi$} = \lambda_2^{-2} \mbox{\boldmath $\Pi$} -
\textstyle{\frac{1}{2}} (\lambda_3^{-2} - \lambda_1^{-2})
    [ (\mathbf{n^+\cdot} \mbox{\boldmath $\Pi$}\mathbf{n^+})
( \mathbf{n^- \cdot}\mbox{\boldmath $\Pi$}\mathbf{n^-})]^{1/2}
    [\mathbf{K^+} \otimes \mathbf{K^-}
        + \mathbf{K^-} \otimes \mathbf{K^+}],
 \end{equation}
where $ \mathbf{K^\pm}$ are unit bivectors given by
\begin{equation}
\mathbf{K^\pm}
= \mbox{\boldmath $\Pi$}\mathbf{n^\pm} / (\mathbf{n^\pm\cdot} \mbox{\boldmath $\Pi$}\mathbf{n^\pm})^{1/2}.
\end{equation}

\vspace{10pt}
 \textbf{Eigenvalues of $\mbox{\boldmath $\chi$}$:}

We note that $\mbox{\boldmath $\chi$}$ is symmetric.
Also $\mbox{\boldmath $\chi$} \mathbf{C^*} = \mathbf{0}$,
so that $\mbox{\boldmath $\chi$}$ has one zero eigenvalue 
corresponding to its eigenbivector $\mathbf{C^*}$.
The other eigenvalues are the roots, $\alpha$, of the quadratic,
\begin{equation}
  \alpha^2 - I_{\mbox{\boldmath $\chi$}}  \alpha
    + II_{\mbox{\boldmath $\chi$}} =0,
\end{equation}
where
\begin{align}
I_{\mbox{\boldmath $\chi$}}
    & = \text{ tr }  \mbox{\boldmath $\chi$}
      = \text{ tr } (\mbox{\boldmath $\Pi$} \mathbb{B}^{-1}
        \mbox{\boldmath$\Pi$})
      = \text{ tr }  \mathbb{B}^{-1} - \mathbf{C^* \cdot}
        \mathbb{B}^{-1} \mathbf{C^*},
\nonumber \\
2 II_{\mbox{\boldmath $\chi$}}
    & = (I_{\mbox{\boldmath $\chi$}})^2 -
          \text{ tr } (\mbox{\boldmath $\chi$}^2)
      = (I_{\mbox{\boldmath $\chi$}})^2
        - \text{ tr } (\mathbb{B}^{-2})
        + 2 \mathbf{C^* \cdot}\mathbb{B}^{-1} \mathbf{C^*}
        - (\mathbf{C^* \cdot} \mathbb{B}^{-1}\mathbf{C^*})^2
\nonumber \\
    & = 2(\mathbf{C^* \cdot} \mathbb{B}^{-1}\mathbf{C^*})/III.
\end{align}

The condition that the quadratic have a double root for $\alpha$ is
that
\begin{equation}
I_{\mbox{\boldmath $\chi$}}^2 = 4II_{\mbox{\boldmath $\chi$}},
\end{equation}
or, equivalently,
\begin{equation}
 III(\text{ tr } \mathbb{B}^{-1} - \mathbf{C^* \cdot} \mathbb{B}^{-1}
\mathbf{C^*})^2 = 4 \mathbf{C^* \cdot} \mathbb{B} \mathbf{C^*}.
\end{equation}
If this is satisfied, the double root of the quadratic is $
\widehat{\alpha}$ (say), given by
\begin{multline} \label{alphaHat}
\widehat{\alpha} = I_{\mbox{\boldmath $\chi$}}/2
             = (\text{ tr }\mathbb{B}^{-1}
          - \mathbf{C^* \cdot} \mathbb{B}^{-1} \mathbf{C^*})/2
\\
= 2(\mathbf{C^* \cdot} \mathbb{B}^{-1} \mathbf{C^*})/III
=\lambda_2^{-2}-\textstyle{\frac{1}{2}}(\lambda_3^{-2}-\lambda_1^{-2})
( \mathbf{n^+ \cdot}\mbox{\boldmath $\Pi$} \mathbf{n^-}),
\end{multline}
on using \eqref{chi3}.

The quadratic has a zero root provided
$II_{\mbox{\boldmath $\chi$}}=0$, or
\begin{equation} \label{C*BC*}
\mathbf{C^* \cdot} \mathbb{B} \mathbf{C^*} = 0,
\end{equation}
in which case $ \mathbf{C^*}$ is any bivector whose ellipse is similar
and similarly situated to a section of the $ \mathbb{B}$-ellipsoid,
other than a central circular section (Throughout this Appendix, it is
assumed that $ \mathbf{C}$ is not an isotropic bivector.)
There is thus an infinity of possible $\mathbf{C^*}$ satisfying
\eqref{C*BC*} -- or equivalently an infinity of possible choices of
$ \mathbf{C^*}$ for which $\mbox{\boldmath $\chi$}$ has two zero
eigenvalues.
In this instance, the third eigenvalue is
$I_{\mbox{\boldmath $\chi$}} = \text{ tr } \mbox{\boldmath $\chi$}$.

The conditions that the quadratic have a double zero root are
$I_{\mbox{\boldmath $\chi$}} =0$,  
$II_{\mbox{\boldmath $\chi$}} =0$, or
\begin{equation} \label{C*PhiC*}
\mathbf{C^* \cdot} \mbox{\boldmath $\Phi$} \mathbf{C^*} = 0, \quad
\mathbf{C^* \cdot} \mathbb{B} \mathbf{C^*} = 0,
\end{equation}
where $\mbox{\boldmath $\Phi$}$ is the real positive definite tensor
given by
\begin{equation}
\mbox{\boldmath $\Phi$} = (\text{tr } \mathbb{B}^{-1}) \mathbf{1} -
\mathbb{B}^{-1}.
\end{equation}
In general, any central plane will cut the ellipsoids
associated with $\mbox{\boldmath $\Phi$}$ and with $\mathbb{B}$ in a
pair of concentric ellipses.
There are two `exceptional' central planes for which each of these
ellipses is similar and similarly situated to the other ellipse 
\cite{BoHa93}.
The two bivectors $\mathbf{C^*}$ which satisfy \eqref{C*PhiC*} may be
obtained by choosing $ \mathbf{C^*}$ to lie in an exceptional central
plane such that its ellipse is similar and similarly situated to the
elliptical section of the $\mbox{\boldmath $\Phi$}$ or $\mathbb{B}$
ellipsoid by the exceptional plane.
In general, there are just two bivectors $\mathbf{C^*}$ which satisfy
\eqref{C*PhiC*}.
Thus, in general, there are just two bivectors $ \mathbf{C^*}$
for which $\mbox{\boldmath $\chi$}$ has a triple zero root.

\vspace{10pt}
 \textbf{Eigenbivectors of $\mbox{\boldmath $\chi$}$:}

How to proceed to determine the eigenbivectors of
$\mbox{\boldmath $\chi$}$ will depend upon whether or not
$\mathbf{n^+ \cdot \mbox{\boldmath $\Pi$} n^+}$ and
$\mathbf{n^- \cdot \mbox{\boldmath $\Pi$} n^-}$ are zero.
Accordingly, we consider the four cases:

Case (i): $\mathbf{n^+ \cdot \mbox{\boldmath $\Pi$} n^+} \ne 0$,
        $\mathbf{n^- \cdot \mbox{\boldmath $\Pi$} n^-} \ne 0$;

Case (iia): $\mathbf{n^+ \cdot \mbox{\boldmath $\Pi$} n^+} = 0$,
        $\mathbf{n^- \cdot \mbox{\boldmath $\Pi$} n^-} \ne 0$;

Case (iib):  $\mathbf{n^+\cdot\mbox{\boldmath $\Pi$} n^+ } \ne 0$,
        $\mathbf{n^- \cdot \mbox{\boldmath $\Pi$} n^-} = 0$;

Case (iii): $\mathbf{n^+\cdot\mbox{\boldmath $\Pi$} n^+ } = 0$,
        $\mathbf{n^- \cdot \mbox{\boldmath $\Pi$} n^-} = 0$.

\vspace{10pt}
 \textit{Case (i):
    $\mathbf{n^+ \cdot \mbox{\boldmath $\Pi$} n^+} \ne 0$,
        $\mathbf{n^- \cdot \mbox{\boldmath $\Pi$} n^-} \ne 0$}.

The form of \eqref{chi4} immediately gives the eigenbivectors of
$\mbox{\boldmath $\chi$}$.
We have
\begin{equation}
 \mbox{\boldmath $\chi$} \mathbf{A^\pm} =\delta^{\pm}\mathbf{A^\pm},
\end{equation}
where
\begin{align} \label{delta+-}
&\delta^{\pm} = \lambda_2^{-2}
 - \textstyle{\frac{1}{2}}(\lambda_3^{-2} - \lambda_1^{-2})
[(\mathbf{n^+ \cdot \mbox{\boldmath $\Pi$} n^+})(\mathbf{n^- \cdot
\mbox{\boldmath $\Pi$} n^-})]^{1/2}
 (\mathbf{K^+ \cdot K^-} \pm 1), \nonumber \\
& \mathbf{A^\pm} =
  (\mathbf{K^+} \pm \mathbf{K^-})/[2(1+ \mathbf{K^+ \cdot K^-})]^{1/2}.
\end{align}
We note that \cite{Dest99b}
\begin{align} \label{A+-properties}
& \mbox{\boldmath $\Pi$} \mathbf{A^\pm} =\mathbf{A^\pm},
\quad
\mathbf{A^\pm \cdot A^\pm} = 1,
\quad
\mathbf{A^+ \cdot A^-} =0,
\nonumber \\
& \mathbf{A^+ \cdot}  \mathbb{B}^{-1} \mathbf{A^-}
    = (\mbox{\boldmath $\Pi$}\mathbf{A^+})
 \mathbf{\cdot}  \mathbb{B}^{-1}(\mbox{\boldmath $\Pi$} \mathbf{A^-})
    = \mathbf{A^+ \cdot}   \mbox{\boldmath $\chi$} \mathbf{A^-}
    = 0.
\end{align}
We also note, on using \eqref{delta+-} and \eqref{A+-properties},
that
\begin{align} \label{delta+-properties}
 \delta^+ - \delta^-  = &
 - (\lambda_3^{-2} - \lambda_1^{-2})
    [(\mathbf{n^+ \cdot \mbox{\boldmath $\Pi$} n^+})
         (\mathbf{n^- \cdot \mbox{\boldmath $\Pi$} n^-})]^{1/2},
\nonumber \\
 \delta^+ + \delta^-  = &
 \mathbf{A^+ \cdot}   \mbox{\boldmath $\chi$} \mathbf{A^+}
 + \mathbf{A^- \cdot}   \mbox{\boldmath $\chi$} \mathbf{A^-} =
  \mathbf{A^+ \cdot}  \mathbb{B}^{-1} \mathbf{A^+}
     +\mathbf{A^- \cdot}  \mathbb{B}^{-1} \mathbf{A^-}
\nonumber \\
& = \text{tr } \mathbb{B}^{-1}
    - \mathbf{C^* \cdot}  \mathbb{B}^{-1} \mathbf{C^*}
  = (II/III) - \mathbf{C^* \cdot}  \mathbb{B}^{-1} \mathbf{C^*},
\\
 \delta^+ \delta^-  = &
 (\mathbf{A^+ \cdot}  \mathbb{B}^{-1} \mathbf{A^+})
     (\mathbf{A^- \cdot}  \mathbb{B}^{-1} \mathbf{A^-})
= (\mathbf{C^* \cdot}  \mathbb{B} \mathbf{C^*})/ III.
\nonumber
\end{align}
Thus, the eigenbivectors of $\mbox{\boldmath $\chi$}$ are
$\mathbf{A^\pm}$, with corresponding distinct eigenvalues
$\delta^\pm$, and $\mathbf{C}$ with corresponding eigenvalue zero.

\vspace{10pt}

\textit{Remark: Special case $\mathbf{C \cdot j} =0$}.

We note that
\begin{multline}
\mbox{\boldmath $\Pi$} \mathbf{n^+} \wedge \mbox{\boldmath $\Pi$}
\mathbf{n^-} = [ \mathbf{C} \wedge ( \mathbf{n^+} \wedge \mathbf{C})]
\wedge [\mathbf{C} \wedge ( \mathbf{n^-} \wedge \mathbf{C})]/(
\mathbf{C \cdot C})^2 \\ = [ (\mathbf{n^+} \wedge \mathbf{n^-})
\mathbf{\cdot} \mathbf{C^*}] \mathbf{C^*} = 2 \alpha \gamma
(\mathbf{C^* \cdot j}) \mathbf{C^*}.
\end{multline}

Assuming that neither $\mbox{\boldmath $\Pi$} \mathbf{n^+}$ nor
$\mbox{\boldmath $\Pi$} \mathbf{n^-}$ is zero, it follows that it is
only when $\mathbf{C^* \cdot j}=0$ that $\mbox{\boldmath $\Pi$}
\mathbf{n^+}$ and $\mbox{\boldmath $\Pi$} \mathbf{n^-}$ are parallel.
Assuming $\mathbf{C^* \cdot j}=0$, then
\begin{equation} \label{specialCase}
\mbox{\boldmath $\Pi$} \mathbf{n^+} = (\alpha C^*_3 - \gamma C^*_1)(
C^*_3 \mathbf{i} - C^*_1 \mathbf{k}), \quad \mbox{\boldmath $\Pi$}
\mathbf{n^-} = (\alpha C^*_3 + \gamma C^*_1)( C^*_3 \mathbf{i} - C^*_1
\mathbf{k}).
\end{equation}

We digress to consider the possibility that $\alpha C^*_3 = \gamma
C^*_1$, in which case $\mbox{\boldmath $\Pi$} \mathbf{n^+} =
\mathbf{0}$. Then, using $C_2^* = 0$, it follows that $\mathbf{C^*} =
C^*_1 \mathbf{n^+}/ \alpha$ so that $\mathbf{C^*}$ is a `real'
bivector, that is, a scalar multiple of a real vector. 
We exclude consideration of this possibility because it leads
to homogeneous waves. 
Similarly, we exclude consideration of the
possibility that $\alpha C^*_3 = - \gamma C^*_1$ so that
$\mbox{\boldmath $\Pi$} \mathbf{n^-} = \mathbf{0}$ and  using $C^*_2 =
0$, leads to $\mathbf{C^*} = C^*_1 \mathbf{n^-}/ \alpha$, again a real
propagation bivector, leading to homogeneous waves.

Returning now to \eqref{specialCase} with $\alpha^2 (C^*_3)^2 \ne
\gamma^2 (C^*_1)^2$, it follows from \eqref{chi4} that
\begin{align}
& \mbox{\boldmath $\chi$} = [\lambda_2^{-2} - (\lambda_3^{-2} -
\lambda_1^{-2})(\alpha^2 (C_3^*)^2 - \gamma^2 (C^*_1)^2)] \mathbf{K^+}
\otimes \mathbf{K^+} +  \lambda_2^{-2} \mathbf{j} \otimes \mathbf{j},
\nonumber \\ & \mathbf{K^+} = \mbox{\boldmath $\Pi$} \mathbf{n^+} /
(\alpha C^*_3 - \gamma C_1^*) = C^*_3 \mathbf{i} - C^*_1 \mathbf{k}.
\end{align}
The eigenbivectors of $ \mbox{\boldmath $\chi$}$ are now 
$\mathbf{C^*}$, $\mathbf{K^+}$, and $\mathbf{j}$, with corresponding
eigenvalues $0$, $\lambda_2^{-2} - (\lambda_3^{-2} -
\lambda_1^{-2})(\alpha^2 (C_3^*)^2 - \gamma^2 (C^*_1)^2)$, and
$\lambda_2^{-2}$.
Because $C_2^* = 0$ and $(C_1^*)^2 + (C_3^*)^2 = 1$, it follows that
equating any two of these eigenvalues will lead to a $\mathbf{C^*}$
which is a real bivector.
So, isotropic eigenbivectors are only possible when the propagation
bivector $\mathbf{C}$ is a real bivector.
Accordingly, only homogeneous circularly-polarized waves are possible
when $C_2^* = 0$.

Thus, this special case $\mathbf{C \cdot j} = 0$ plays no role in the
study of circularly-polarized inhomogeneous plane waves.

\vspace{10pt}
 \textit{Case (iia):
    $\mathbf{n^+ \cdot \mbox{\boldmath $\Pi$} n^+} = 0$,
        $\mathbf{n^- \cdot \mbox{\boldmath $\Pi$} n^-} \ne 0$}.

In this case, 
$(\mbox{\boldmath $\Pi$} \mathbf{n^+}) 
	\cdot (\mbox{\boldmath $\Pi$} \mathbf{n^+}) =0$, so that
$\mbox{\boldmath $\Pi$} \mathbf{n^+}$ is an isotropic bivector.
Indeed, the orthogonal projection of the ellipse of $\mathbf{C^*}$ 
onto the plane with normal $\mathbf{n^+}$ is a circle.
So
\begin{equation} 
\mbox{\boldmath $\chi$} \mbox{\boldmath $\Pi$} \mathbf{n^+}
 = \nu_\perp \mbox{\boldmath $\Pi$} \mathbf{n^+}, \quad
\nu_\perp = \lambda_2^{-2} - 
	\textstyle{\frac{1}{2}}(\lambda_3^{-2} - \lambda_1^{-2})
 \mathbf{n^+ \cdot \mbox{\boldmath $\Pi$} n^-}.
\end{equation}

We also have 
$\mbox{\boldmath $\chi$}  \mathbf{C} = \mathbf{0}$, 
$\mbox{\boldmath $\Pi$} \mathbf{C} = \mathbf{0}$.
It may be checked that there is no bivector 
$\mbox{\boldmath $\Pi$}  \mathbf{n^+}
	 +\epsilon \mbox{\boldmath $\Pi$} \mathbf{n^-}$
($\epsilon \ne 0$) which is orthogonal to the eigenbivector
$\mathbf{C}$ and forming with 
$\mbox{\boldmath $\Pi$}  \mathbf{n^+}$ a third
eigenbivector of $\mbox{\boldmath $\chi$}$.
Thus, $\mbox{\boldmath $\Pi$} \mathbf{n^+}$
is an isotropic eigenbivector of $\mbox{\boldmath $\chi$}$ with
eigenvalue $\nu_\perp$.

\vspace{10pt}
 \textit{Case (iib):
    $\mathbf{n^+ \cdot \mbox{\boldmath $\Pi$} n^+} \ne 0$,
        $\mathbf{n^- \cdot \mbox{\boldmath $\Pi$} n^-}= 0$}.

This is similar to Case (iia).
In this case, $\mbox{\boldmath $\Pi$} \mathbf{n^-}$
is an isotropic eigenbivector of $\mbox{\boldmath $\chi$}$ with
eigenvalue $\nu_\perp$.
Also, of course, $\mbox{\boldmath $\Pi$} \mathbf{n^-}$ is orthogonal 
to $\mathbf{C}$, the eigenbivector with eigenvalue zero.

\vspace{10pt}
 \textit{Case (iii):
    $\mathbf{n^+ \cdot \mbox{\boldmath $\Pi$} n^+} = 0$,
        $\mathbf{n^- \cdot \mbox{\boldmath $\Pi$} n^-}= 0$}.

Here both $\mbox{\boldmath $\Pi$} \mathbf{n^+}$ and 
$\mbox{\boldmath $\Pi$} \mathbf{n^-}$ are isotropic bivectors -- 
the ellipse of $\mathbf{C^*}$  when projected upon a plane with normal
 $\mathbf{n^+}$ is a circle, and so is its projection onto a plane 
with normal $\mathbf{n^-}$.
Also,  $\mathbf{n^+ \cdot \mbox{\boldmath $\Pi$} n^+} = 0$,
        $\mathbf{n^- \cdot \mbox{\boldmath $\Pi$} n^-}= 0$ lead to
\begin{equation} \label{Aiii}
\mathbf{C^*} \cdot ( \mathbf{n^+ \pm n^-} )  =0, \quad
(\mathbf{C^* \cdot n^\pm})^2 =1,
\end{equation}
so that the ellipse of $\mathbf{C^*}$ lies \textbf{either} 
in a plane with normal along $\mathbf{n^+} + \mathbf{n^-}$ 
($= 2 \alpha \mathbf{i}$, recall \eqref{n+-}), 
which is along the internal bisector of the angle between 
$\mathbf{n^+}$ and $\mathbf{n^-}$,
\textbf{or} in a plane with normal along $\mathbf{n^+} -\mathbf{n^-}$ 
($= 2 \gamma \mathbf{k}$, recall \eqref{n+-}), 
which is along the external bisector of the angle between 
$\mathbf{n^+}$ and $\mathbf{n^-}$.
Thus, there are only two cases:
\begin{equation} 
\text{(a) } \mathbf{C^* \cdot i} = 0, \quad
\text{(b) } \mathbf{C^* \cdot k} = 0.
\end{equation}

In case (a), we find (recall \eqref{n+-}) from \eqref{Aiii},
\begin{equation} 
\mathbf{C^*} = ( \pm i \alpha \mathbf{j} + \mathbf{k})/ \gamma,
\quad
\mathbf{n^+ \cdot \mbox{\boldmath $\Pi$} n^-} = 2 \alpha^2,
\quad
\mbox{\boldmath $\chi$} \mbox{\boldmath $\Pi$} \mathbf{n^\pm} 
 = \lambda_1^{-2} \mbox{\boldmath $\Pi$} \mathbf{n^\pm}.
\end{equation}
Also, these isotropic eigenbivectors 
$\mbox{\boldmath $\Pi$} \mathbf{n^+}$ and 
$\mbox{\boldmath $\Pi$} \mathbf{n^-}$ are linearly independent.
Indeed, if $\mbox{\boldmath $\Pi$} \mathbf{n^+} 
 = \mu \mbox{\boldmath $\Pi$} \mathbf{n^-}$ for some scalar $\mu$,
then $0 = \mathbf{n^+ \cdot} \mbox{\boldmath $\Pi$} \mathbf{n^+}
 = \mu \mathbf{n^+ \cdot} \mbox{\boldmath $\Pi$} \mathbf{n^-} = 
 2 \mu \alpha^2$, so that $\mu = 0$.

In case (b), we find
\begin{equation} 
\mathbf{C^*} = ( \mathbf{i} \pm i \gamma \mathbf{j})/ \alpha,
\quad
\mathbf{n^+ \cdot \mbox{\boldmath $\Pi$} n^-} = 2 \gamma^2,
\quad
\mbox{\boldmath $\chi$} \mbox{\boldmath $\Pi$} \mathbf{n^\pm} 
 = \lambda_3^{-2} \mbox{\boldmath $\Pi$} \mathbf{n^\pm}.
\end{equation}
As before, the isotropic eigenbivectors of $\mbox{\boldmath $\chi$}$,
$\mbox{\boldmath $\Pi$} \mathbf{n^+}$ and 
$\mbox{\boldmath $\Pi$} \mathbf{n^-}$ are linearly independent.

Thus, $\mbox{\boldmath $\chi$}$ has eigenvalues zero and double
eigenvalue $\lambda_1^{-2}$ (in case (a)), or $\lambda_3^{-2}$ 
(in case (b)), and corresponding eigenbivector $\mathbf{C}$ and 
isotropic eigenbivectors $\mbox{\boldmath $\Pi$} \mathbf{n^+}$ and 
$\mbox{\boldmath $\Pi$} \mathbf{n^-}$.
Of course, $\mathbf{C}$, $\mbox{\boldmath $\Pi$} \mathbf{n^+}$, and 
$\mbox{\boldmath $\Pi$} \mathbf{n^-}$ are linearly independent.
Indeed, if for some scalars $p$, $q$,
\begin{equation}
\mathbf{C^*} = p \mbox{\boldmath $\Pi$} \mathbf{n^+} 
	+ q \mbox{\boldmath $\Pi$} \mathbf{n^-},
\end{equation}
then
\begin{equation} 
\mathbf{0} = \mbox{\boldmath $\Pi$} \mathbf{C^*} 
 = p \mbox{\boldmath $\Pi$} \mathbf{n^+} 
	+ q \mbox{\boldmath $\Pi$} \mathbf{n^-},
\end{equation}
so that $\mbox{\boldmath $\Pi$} \mathbf{n^+}$ is a scalar multiple
of  $\mbox{\boldmath $\Pi$} \mathbf{n^-}$.
But we have already seen that 
$\mbox{\boldmath $\Pi$} \mathbf{n^+}$ and 
$\mbox{\boldmath $\Pi$} \mathbf{n^-}$ are linearly independent.
Thus $\mathbf{C}$, $\mbox{\boldmath $\Pi$} \mathbf{n^+}$, and 
$\mbox{\boldmath $\Pi$} \mathbf{n^-}$ are linearly independent.

\end{document}